\newcommand\apjcls{1}
\newcommand\aastexcls{2}
\newcommand\othercls{3}

\newcommand\papercls{\aastexcls}
\documentclass[tighten, times, twocolumn]{aastex62}  

\if\papercls \apjcls
\usepackage{apjfonts}
\else\if\papercls \othercls
\usepackage{epsfig}
\usepackage{margin}
\usepackage{times}
\fi\fi
\usepackage{ifthen}
\usepackage{natbib}
\usepackage{amssymb, amsmath}
\usepackage{appendix}
\usepackage{etoolbox}
\usepackage[T1]{fontenc}
\usepackage{paralist}

\if\papercls \apjcls
\newcommand\aas{\ref@jnl{AAS Meeting Abstracts}}
\newcommand\dps{\ref@jnl{AAS/DPS Meeting Abstracts}}
\newcommand\maps{\ref@jnl{MAPS}}
\else\if\papercls \othercls
\usepackage{astjnlabbrev-jh}
\fi\fi

\if\papercls \aastexcls
\hypersetup{citecolor=blue, 
            linkcolor=blue, 
            menucolor=blue, 
            urlcolor=blue}  
\else
\usepackage[
bookmarks=true,           
bookmarksnumbered=true,   
colorlinks=true,          
citecolor=blue,           
linkcolor=blue,           
menucolor=blue,           
urlcolor=blue,            
linkbordercolor={0 0 1},  
pdfborder={0 0 1},
frenchlinks=true]{hyperref}
\fi

\if\papercls \othercls

\else

\fi

\providecommand{\adsurl}[1]{\href{#1}{ADS}}

\makeatletter
\patchcmd{\NAT@citex}
  {\@citea\NAT@hyper@{%
     \NAT@nmfmt{\NAT@nm}%
     \hyper@natlinkbreak{\NAT@aysep\NAT@spacechar}{\@citeb\@extra@b@citeb}%
     \NAT@date}}
  {\@citea\NAT@nmfmt{\NAT@nm}%
   \NAT@aysep\NAT@spacechar\NAT@hyper@{\NAT@date}}{}{}

\patchcmd{\NAT@citex}
  {\@citea\NAT@hyper@{%
     \NAT@nmfmt{\NAT@nm}%
     \hyper@natlinkbreak{\NAT@spacechar\NAT@@open\if*#1*\else#1\NAT@spacechar\fi}%
       {\@citeb\@extra@b@citeb}%
     \NAT@date}}
  {\@citea\NAT@nmfmt{\NAT@nm}%
   \NAT@spacechar\NAT@@open\if*#1*\else#1\NAT@spacechar\fi\NAT@hyper@{\NAT@date}}
  {}{}
\makeatother

\makeatletter
\DeclareRobustCommand{\lowcase}[1]{\@lowcase#1\@nil}
\def\@lowcase#1\@nil{\if\relax#1\relax\else\MakeLowercase{#1}\fi}
\makeatother

\DeclareSymbolFont{UPM}{U}{eur}{m}{n}
\DeclareMathSymbol{\umu}{0}{UPM}{"16}
\let\oldumu=\umu
\renewcommand\umu{\ifmmode\oldumu\else\math{\oldumu}\fi}

\if\papercls \othercls

\else

\fi

\let\oldsim=\sim
\renewcommand\sim{\ifmmode\oldsim\else\math{\oldsim}\fi}
\let\oldpm=\pm
\renewcommand\pm{\ifmmode\oldpm\else\math{\oldpm}\fi}
\newcommand\by{\ifmmode\times\else\math{\times}\fi}

\newbox{\wdbox}
\renewcommand\c{\setbox\wdbox=\hbox{,}\hspace{\wd\wdbox}}
\renewcommand\i{\setbox\wdbox=\hbox{i}\hspace{\wd\wdbox}}

\newcount\timect
\newcount\hourct
\newcount\minct
\newcommand\now{\timect=\time \divide\timect by 60
         \hourct=\timect \multiply\hourct by 60
         \minct=\time \advance\minct by -\hourct
         \number\timect:\ifnum \minct < 10 0\fi\number\minct}

\catcode`@=11

\newcommand\comment[1]{}

\newcommand\commenton{\catcode`\%=14}

\renewcommand\math[1]{$#1$}
\newcommand\mathshifton{\catcode`\$=3}

\let\atab=&
\newcommand\atabon{\catcode`\&=4}

\let\oldmsp=\sp
\let\oldmsb=\sb
\def\sp#1{\ifmmode
           \oldmsp{#1}%
         \else\strut\raise.85ex\hbox{\scriptsize #1}\fi}
\def\sb#1{\ifmmode
           \oldmsb{#1}%
         \else\strut\raise-.54ex\hbox{\scriptsize #1}\fi}
\newbox\@sp
\newbox\@sb
\def\sbp#1#2{\ifmmode%
           \oldmsb{#1}\oldmsp{#2}%
         \else
           \setbox\@sb=\hbox{\sb{#1}}%
           \setbox\@sp=\hbox{\sp{#2}}%
           \rlap{\copy\@sb}\copy\@sp
           \ifdim \wd\@sb >\wd\@sp
             \hskip -\wd\@sp \hskip \wd\@sb
           \fi
        \fi}
\def\msp#1{\ifmmode
           \oldmsp{#1}
         \else \math{\oldmsp{#1}}\fi}
\def\msb#1{\ifmmode
           \oldmsb{#1}
         \else \math{\oldmsb{#1}}\fi}

\def\supon{\catcode`\^=7}

\def\subon{\catcode`\_=8}

\def\supsubon{\supon \subon}

\newcommand\actcharon{\catcode`\~=13}

\newcommand\paramon{\catcode`\#=6}

\comment{And now to turn us totally on and off...}

\newcommand\reservedcharson{ \commenton  \mathshifton  \atabon  \supsubon 
                             \actcharon  \paramon}

\catcode`@=12
\reservedcharson

\if\papercls \apjcls

\else

\fi

\usepackage{xspace}
\usepackage{amsmath}
\usepackage{hyperref}
\usepackage[figuresright]{rotating}
\usepackage{enumerate}
\usepackage{bm}
\usepackage{csquotes}
\MakeOuterQuote{"}

\newcommand{\nh}{$N(\text{H})$\xspace}

\newcommand{\alam}{$A_\lambda$\xspace}

\newcommand{\ebv}{$E(B-V)$\xspace}

\newcommand{\redchisq}{$\chi^2_\text{red}$\xspace}
\newcommand{\bnir}{$\beta_\text{NIR}$\xspace}

\newcommand{\um}{$\mu$m\xspace}

\newcommand{\Rv}{$R_V$\xspace}

\newcommand{\teff}{$T_\text{eff}$\xspace}
\newcommand{\logg}{$\log g$\xspace}
\newcommand{\feh}{[Fe/H]\xspace}

\newcommand{\Elami}{$E(\lambda-i)$\xspace}

\newcommand{\lamr}{$\lambda-r$\xspace}

\newcommand{\Euz}{$E(u-z)$\xspace}
\newcommand{\invlam}{$\lambda^{-1}$\xspace}

\newcommand{\sfdebv}{$E(B-V)_\text{SFD}$\xspace}
\newcommand{\stdresid}{$\sigma_\text{res}$\xspace}
\newcommand{\ai}{$A_i$\xspace}

\newcommand{\Rvstd}{$R_V=3.1$\xspace}
\newcommand{\bnirstd}{$\beta_\text{NIR}=1.84$\xspace}
\newcommand{\EBVcutlo}{$E(B-V)_\text{SFD}<0.0125$\xspace}
\newcommand{\EBVcuthi}{$E(B-V)_\text{SFD}>0.0125$\xspace}
\newcommand{\Euzcut}{$E(u-z)>1$\xspace}

\newcommand{\hub}{$H_0$\xspace}

\newcommand{\RH}{$R_H$\xspace}
\newcommand{\AH}{$A_H$\xspace}
\newcommand{\RHstd}{$R_H=0.386$\xspace}
\newcommand{\RHus}{$R_H=0.345$\xspace}
\newcommand{\RHuserr}{$R_H=0.345\pm0.007$\xspace}
\newcommand{\Rvhub}{$R_V=3.3$\xspace}
\newcommand{\EVI}{$E(V-I)$\xspace}

\newcommand{\Eri}{$E(r-i)$\xspace}

\newcommand{\sigbeta}{$\sigma(\beta_\text{NIR})$\xspace}
\newcommand{\sigRv}{$\sigma(R_V)$\xspace}

\newcommand{\betamath}{\beta_\text{NIR}}

\newcommand{\Hhub}{F160W\xspace}
\newcommand{\Vhub}{F555W\xspace}
\newcommand{\Ihub}{F814W\xspace}

\newcommand{\allstars}{56,649\xspace}
\newcommand{\hilatstars}{47,296\xspace}
\newcommand{\lolatstars}{9,353\xspace}

\newcommand{\survtype}{survey+subclass\xspace}

\defcitealias{CCM}{CCM}
\newcommand{\ctccm}{\citetalias{CCM}\xspace}

\defcitealias{Wang2019}{WC19}

\defcitealias{Schlegel1998}{SFD}
\newcommand{\ctsfd}{\citetalias{Schlegel1998}\xspace}

\shorttitle{The Near-Infrared Extinction Curve}
\shortauthors{Butler \& Salim}

\graphicspath{{./}{figures/}}

\begin{document}

\title{The Near-Infrared Extinction Law at High and Low Galactic Latitudes}

\correspondingauthor{Robert E. Butler}
\email{bbutler3@iu.edu}

\author[0000-0003-2789-3817]{Robert E. Butler}
\affiliation{Department of Astronomy, Indiana University, 727 E. Third Street, Bloomington, IN 47405}

\author[0000-0003-2342-7501]{Samir Salim}
\affiliation{Department of Astronomy, Indiana University, 727 E. Third Street, Bloomington, IN 47405}


\begin{abstract}

The Milky Way extinction curve in the near-infrared (NIR) follows a power law form, but the value of the slope, $\beta_\text{NIR}$, is debated. Systematic variations in the slope of the Milky Way UV extinction curve are known to be correlated with variations in the optical slope (through $R_V$), but whether such a dependence extends to the NIR is unclear. Finally, because of low dust column densities, the NIR extinction law is essentially unconstrained at high Galactic latitudes where most extragalactic work takes place.
In this paper, we construct extinction curves from 56,649 stars with SDSS and 2MASS photometry, based on stellar parameters from SDSS spectra. We use dust maps to identify dust-free stars, from which we calibrate the relation between stellar parameters and intrinsic colors. Furthermore, to probe the low-dust regime at high latitudes, we use aggregate curves based on many stars.
We find no significant variation of $\beta_\text{NIR}$ across low-to-moderate dust columns ($0.02<E(B-V)\lesssim1$), and report average $\beta_\text{NIR}=1.85\pm0.01$, in agreement with \citet{Fitzpatrick2019}, but steeper than \citet{CCM} and \citet{Fitzpatrick1999}. We also find no intrinsic correlation between $\beta_\text{NIR}$ and \Rv (there is an apparent correlation which is the result of the correlated uncertainties in the two values). These results hold for typical sightlines; we do not probe very dusty regions near the Galactic Center, nor rare sightlines with $R_V>4$. Finally, we find $R_H=0.345\pm0.007$ and comment on its bearing on Cepheid calibrations and the determination of $H_0$.

\end{abstract}



\section{Introduction} \label{sec:intro}

Determinations of the dust extinction law in our galaxy, and its variations along different lines of sight, are relevant for many areas of astronomy (see the recent reviews of \citealp{Salim2020} and \citealp{Hensley2021}). The result from \citet[][hereafter \ctccm]{CCM} that UV and optical extinction curves in the Milky Way belong to a single-parameter family was a watershed development toward systematizing observed variations. That single parameter, \Rv, is essentially the slope of the extinction curve in the optical region ($1/R_V = A_B/A_V -1$). However, in \ctccm, this \Rv dependence did not extend to the near-infrared (NIR) part of the extinction curve, where
they fit a power law of the form $A_\lambda \propto \lambda^{-\betamath}$ to the data from \cite{Rieke1985}, yielding $\betamath=1.61$. \ctccm find this slope is applicable regardless of \Rv. Indeed, the existence of such a "universal" NIR extinction law had been found in several prior and contemporary studies \citep[e.g.,][]{Jones1980, Koornneef1983, Smith1987}. 

Though the power-law nature of the NIR extinction curve has been concretely established by numerous studies, the value of the slope varies from study to study. Early studies prior to and after \citet{Rieke1985} found values in the range between about 1.6 to 1.85 (see the reviews of \citealp{Draine1989}, \citealp{Mathis1990}, and \citealp{Draine2003}). \citet{Martin1990} found one of the highest values of the time, $\betamath=1.84$; this value was subsequently used in the parameterization developed by \cite{fm07}. More recent studies of the NIR portion of the curve have found higher values for \bnir \citep[up to 2.64;][]{Gosling2009}. 
While there is clearly a wide spread in \bnir determined between studies, the origin of the range is unclear. One possible explanation is variance with the environment or amount of dust along sightlines, but this has not been concretely shown. 
While different studies have found differing slopes, few studies claimed that there is a systematic change in this slope. The general assumption still is that it is universal, and evidence for dependence on \Rv has been limited to very small-scale studies \citep{Fitzpatrick2009, Decleir2022}.
Many of the recent studies finding high \bnir have focused on the extreme-extinction region in the Galactic Center \citep[e.g.,][]{Nishiyama2009, Gosling2009, Schodel2010, Hosek2018, Nogueras-Lara2021, Sanders2022}, which, if the curve is dependent on the environment, could explain higher values at least in part. 

What essentially all previous studies of NIR extinction laws have had in common is their focus in the plane of the Galaxy, i.e., low Galactic latitude regions of relatively high dust content. It is inherently more difficult to study dust in regions where there is very little, especially in the NIR. The effects of the dust on incoming light are subtle and the photometry must therefore be very precise. 
An additional problem for the determination of dust extinction curves at high latitudes arises for the empirical pair method \citep[e.g.,][]{Fitzpatrick1990} because the comparison stars rarely have zero dust along their sightlines and may have a comparable amount of dust to target high-latitude sightlines. This is not an issue where models are used to constrain extinction \citep[see, e.g.,][]{Fitzpatrick2005}, but the requirement of high accuracy remains.

Although it is generally assumed that the nature of the dust (and therefore the extinction curve) is the same at high latitudes as it is at low latitudes, this has not been confirmed in the NIR. Since extragalactic astronomy and cosmology is largely conducted at high latitudes, it is important to constrain extinction in that low-dust regime, especially in the era of JWST \citep{Gardner2006} and LSST \citep{Ivezic2019}, in addition to the upcoming Euclid \citep{Racca2016} and Nancy Grace Roman \citep{Akeson2019} missions, which will require extinction corrections of high accuracy. 

 Among the most prominent open questions in astronomy and cosmology is the accurate value of the Hubble constant (\hub). The most recent result based on Cepheids and Type Ia supernovae \citep{Riess2022} is in $5\sigma$ tension with the result derived from cosmic microwave background anisotropies \citep{Planck2020}. The local determination of \hub uses Cepheid variables, which follow a tight relation between period and luminosity \citep{Leavitt1912}, allowing for very precise distance measurements. The observed magnitudes of the Cepheids must be corrected for dust extinction along the sightlines. There have been extensive studies into Cepheid calibrations and SN host galaxy dust, with no small amount of attention being paid to the dust extinction \citep[e.g., ][]{Riess2016, Riess2021, Mandel2017, Follin2018, Brout2021, Meldorf2023}. Precise NIR extinction values are crucial for Cepheid calibrations specifically (the first rung in the distance ladder), and knowledge of the appropriate NIR extinction law is required. 

In recent years, there has been a proliferation of extinction studies which use large samples of stars taken from surveys \citep[e.g.,][]{Peek2010, Schlafly2010, Schlafly2011, Berry2012}. Given such a sample, one can apply several different methods to derive intrinsic colors for stars, which can then be subtracted from observed colors to obtain reddenings. 
Some of these methods employ template spectra or atmospheric models to generate simulated comparison stars \citep[e.g.,][]{Jones2011, Schlafly2011}, while others use a set of stars with very low extinction to construct a relation between their photometry and stellar parameters, usually \teff, \logg, and \feh \citep[e.g.,][]{Yuan2013,Yuan2015,Li2018,Sun2018,Wang2019}. 
We follow a procedure of the latter kind to determine intrinsic colors for our sample of stars, with the goal of investigating NIR extinction curves at both high and low latitudes.

In this study, we attempt to clarify the potential dependence of \bnir on the characteristics of individual sightlines. Namely, we look for a correlation between \bnir and the dust column density along a sightline, and separately for a correlation between \bnir and \Rv. 
We construct extinction curves across a wide range of dust columns, made possible specifically at high latitudes by a large sample size. 
In the low-dust (high-latitude) regime, NIR extinction is most poorly constrained, simply due to the very small effect of dust on the incoming light. Because of this, we use aggregates of large numbers of stars from optical and NIR surveys (SDSS, 2MASS) to increase our sensitivity to the effects of the dust. 

Section \ref{sec:data} describes our optical and NIR data sources. Section \ref{sec:methods} details the methods we use for deriving intrinsic colors and subsequently constructing extinction curves. In Section \ref{sec:results}, we present the results of our study, relating \bnir to the dust column and also to \Rv. We additionally comment on the bearing of our results on the Cepheid distance ladder calibration in the pursuit of a precise determination of the Hubble constant, \hub. In Section \ref{sec:discussion}, we discuss our results in the context of prior results from the literature.


\section{Data and Sample} \label{sec:data}

We incorporate both optical and near-IR photometry of Galactic stars into our analysis, along with stellar parameters derived from optical spectra. 

\subsection{Data Sources} \label{subsec:datasources}

We draw our optical photometry (\textit{ugriz}) and spectroscopy from the Sloan Digital Sky Survey \citep[SDSS;][]{York2000} Data Release 16 \citep{Ahumada2020}, in particular its Legacy (henceforth referred to as SDSS) and SEGUE components, which collectively cover 14,555 square degrees. We use PSF magnitudes, which are most appropriate for stars.

The spectra from which our stellar parameters are derived come from any of three separate (but closely related) spectroscopic campaigns: the SDSS Legacy Survey (reported in the database as \texttt{SDSS}), SEGUE-1 \citep{Yanny2009}, and SEGUE-2 \citep{Eisenstein2011}. These surveys all employed the same twin spectrographs with wavelength coverage from 3900 \AA\ to 9000 \AA\ at a moderate resolution of $R\sim1800$.
We take the stellar parameters \teff, \logg, and \feh as reported in the DR16 sppParams table, derived using the SEGUE Stellar Parameter Pipeline \citep[SSPP;][]{Lee2008}. 
\citet{Lee2008} derive these three parameters using multiple methods, with the weighted averages reported as \texttt{adop}. The color-independent, spectroscopically-determined weighted averages are reported as \texttt{spec}. 
We choose the \texttt{spec} parameters instead of the recommended \texttt{adop} parameters because the former are independent from photometry. We also take the spectroscopic \texttt{subClass} of the stars.

For \textit{JHK} near-infrared photometry, we use the Two Micron All Sky Survey \citep[2MASS;][]{Skrutskie2006} Point Source Catalog (PSC), which covers the SDSS footprint in its entirety. We use the "default" magnitudes as reported in the PSC. Note that we will use $K$ to refer to the 2MASS $K_s$ band henceforth.
We also use UKIDSS near-infrared photometry \citep{Lawrence2007}, which covers only a portion of the SDSS footprint (but is deeper than 2MASS) to perform checks on 2MASS.

\subsection{Sample Selection} \label{subsec:sample}

We queried SDSS within the DR16 context in CasJobs to obtain the base optical sample from SDSS. We require primary objects classified as stars (\texttt{class=`STAR'}) and with a clean photometry flag (\texttt{clean=1}). 
We also require that valid \teff, \logg, and \feh values are present, and 
that the spectroscopic flag \texttt{zWarning} is set to 0 or 16, as recommended by SDSS documentation. 

From this original optical sample of $\sim$342,000 stars with spectra, we make several further quality cuts on the photometric data. We make a cut on the quality of field (\texttt{score}$<0.7$), 
which removes a small clump of poor-quality fields. 
Finally, we require that magnitudes in all SDSS bands are present for all stars, and that their photometric uncertainties are less than $0.1$ mag. After these cuts, our base optical sample contains $\sim$314,000 stars. 

We make use of the \texttt{TwoMass} table within CasJobs to get \textit{JHK} photometry for our sources. Since 2MASS is less deep than SDSS, the stars for which we recover \textit{JHK} photometry number about 60\% of the optical sample. 
Comparing 2MASS and UKIDSS, we find that 2MASS tends to overestimate the brightness of an object at faint magnitudes in \textit{K} band. 
To eliminate this issue, we remove all stars with $K>14.8$ from all analyses that pertain to the near-IR. 

In a large-survey-based endeavor such as the current study, it may not be possible to eliminate all photometric anomalies using only catalog flags and simple cuts. 
We therefore inspect color-color plots in order to eliminate any remaining outliers. In the optical, we find $i$ to be the most potentially problematic band, so we eliminate outliers
beyond 6 robust $\sigma$ away from the running median in a plot of $r-i$ vs. $g-z$. We also find it necessary to remove any stars with $(u-g)<0.7$. 
We make an additional cut 
determined using a similar method to the $r-i$ cut above, but this time using a plot of $r-J$ vs. $g-z$. We have verified that our outlier cuts are not removing any real measurements arising from atypical $R_V$ values.

\begin{figure}[t!]
    \centering
    \includegraphics[width=\linewidth,keepaspectratio]{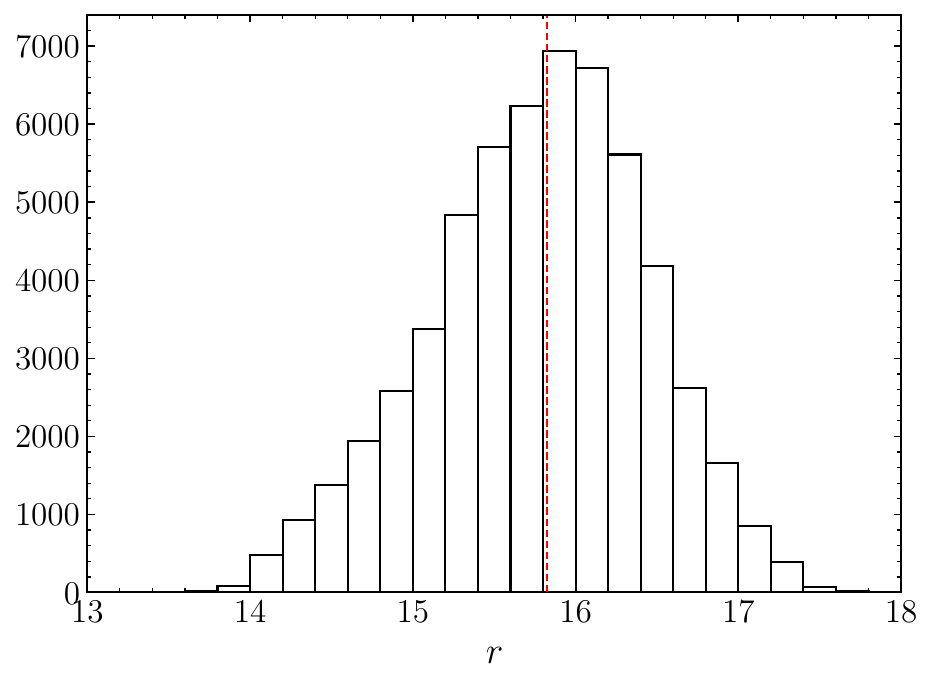}
    \caption{Distribution of observed SDSS $r$ magnitudes for the \allstars stars used in our final program sample for the main analysis. The dashed red vertical line is the median, $r=15.8$.} 
    \label{fig:rmag}
\end{figure}

For the calibration sample (see Section \ref{sec:methods}), we use only stars with \EBVcutlo. In addition, we do not require that the stars have valid measurements in \textit{JHK} (as we do for the program sample); rather, they only need valid measurements in $r$ and the band in question for calibration. This allows us to maximize our calibration sample to allow for better precision in intrinsic color estimates. All other cuts still apply for this sample.

The final (optical+NIR) sample from which we draw reddened (program) stars for this study contains \allstars.
Figure \ref{fig:rmag} shows the distribution of $r$ magnitudes for the stars in our final program sample, along with their median. The distribution is perhaps more symmetrical than expected for a general selection of Milky Way stars; this is likely due to the matching with 2MASS, which removes many of the fainter stars. Since we are not concerned with completeness in this study, these effects are not important. 

\begin{figure}[t!]
    \centering
    \includegraphics[width=\linewidth,keepaspectratio]{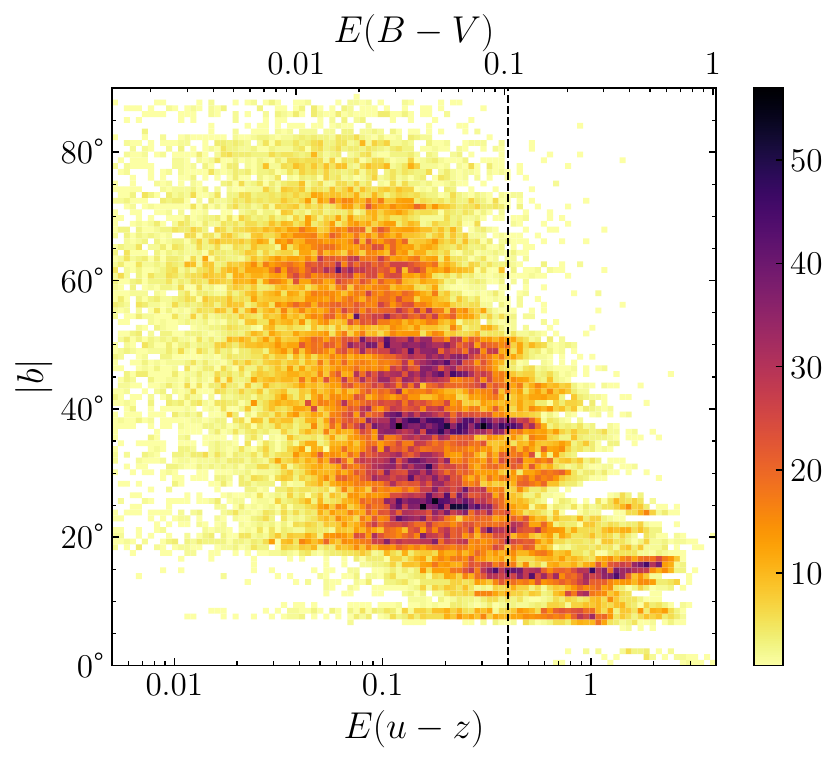}
    \caption{A 2D histogram of the absolute value of Galactic latitude vs. \Euz, a proxy for dust. All of the \allstars stars used in our final program sample for the main analysis are included, with the exception of a small number which have very low \Euz. The dashed vertical line indicates the approximate separation between high-latitude (low-dust; left) and low-latitude (high-dust; right) regimes. We also provide a secondary $x$-axis to show the more familiar $E(B-V)$ value, determined via a simple ratio from \Euz (see Section \ref{subsec:extcurves}).} 
    \label{fig:lat}
\end{figure}

Figure \ref{fig:lat} characterizes our sample with respect to the absolute value of Galactic latitude and the log of reddening \Euz. We consider \Euz a proxy for the dust column (see Section \ref{subsec:colex}). We cover a wide range of absolute latitudes, but note that very few stars lie below $|b|\approx 7^\circ$. There is a shift in the latitude distribution around $E(u-z)=0.4$ (or equivalently $E(B-V)=0.1$), which we will use to distinguish between dust columns typical of low and high latitude. 
The banding in the figure is a result of the SEGUE discrete data-taking pattern.

The numbers of stars in each spectral subclass from our final program sample are shown in Figure \ref{fig:spclass}. The relative representation of each of the three constituent spectroscopic campaigns is also shown. Most of our program stars come from SEGUE, and fall generally within the F type.

\section{Methods} \label{sec:methods}

In this study, we endeavor to obtain parameterized near-infrared extinction curves 
across a wide range of dust column densities. To do this, we need (1) a way to estimate intrinsic colors, (2) color excesses (reddenings) calculated using (1), and (3) a model to which the color excesses can be fitted to obtain a parametric absolute extinction curve ($A(\lambda)$). These steps are presented in the sections to follow.

\subsection{Intrinsic Color Estimates} \label{subsec:regression}

\begin{figure}[t!]
    \centering
    \includegraphics[width=\linewidth,keepaspectratio]{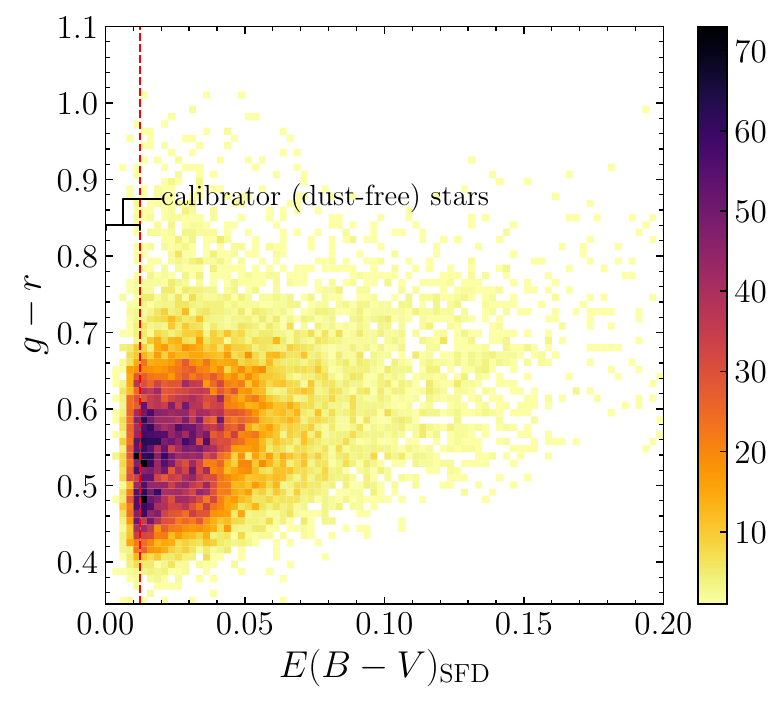}
    \caption{2D histogram of observed SDSS $g-r$ color vs. $E(B-V)$ from \cite{Schlegel1998} for SEGUE2 F9 stars in our original sample. The stars to the left of the red line indicating $E(B-V)<0.0125$ (essentially unreddened stars) were used to determine the intrinsic color calibration by fitting to Equation \ref{eqn:regression}.
    } 
    \label{fig:g-rEBV}
\end{figure}

\begin{figure}[t!]
    \centering
    \includegraphics[width=\linewidth,keepaspectratio]{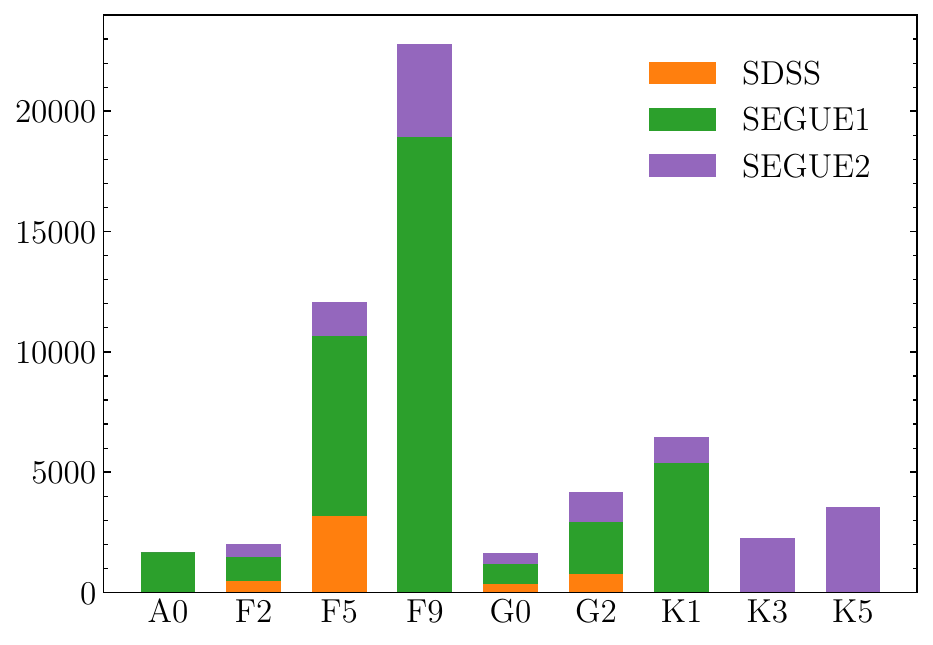}
    \caption{Distribution of SDSS spectral subclasses (\texttt{SUBCLASS}) for the \allstars stars used in our final program sample for the main analysis. The stacked colored bars indicate the relative numbers of stars in each spectral subclass which were measured by the SDSS (Legacy), SEGUE1, and SEGUE2 spectroscopic campaigns.} 
    \label{fig:spclass}
\end{figure}

In order to construct an extinction curve for the dust along a given sightline, estimates of the intrinsic (unreddened) colors of the object are required. 
We make use of stellar parameters determined from SDSS spectroscopy
to estimate the intrinsic colors of stars in our sample. In general terms, the spectral type of a star is related to its color, but there is still a significant spread of colors even within one subclass. 
We show an example \survtype combination (SEGUE2 F9) in Figure \ref{fig:g-rEBV}. 
There is a general trend such that the observed colors become redder with increasing \sfdebv, the reddening as determined by \citet[][SFD]{Schlegel1998}. 
However, there is quite a large range ($\sim$0.25) in observed $g-r$ color, even near $E(B-V)_\text{SFD}=0$, i.e., where the \ctsfd dust map tells us there is no dust. Thus, as expected, a pair-matching technique for constructing extinction curves, based on spectral type alone, is rather imprecise for later spectral types.
Fortunately, with stellar parameters (\logg, \feh, and \teff) in addition to the spectral class, we can more closely relate a star's intrinsic color to its physical properties than if spectral type alone were used. 

In order to calibrate the relation between the intrinsic color and stellar parameters (including spectral type), we select the stars from our sample with very low reddening, \EBVcutlo.
Note that the \sfdebv value is the upper limit because the reddening map includes measures the full column of dust in the direction of the star, and the star may not have all of this dust in front of it.
The stars are then grouped by spectral subclass and survey (e.g., a group may be SEGUE2 F9). In principle, grouping by survey should not be necessary, but we maintain it to remove any potential systematics.


Within each \survtype grouping,  
we carry out calibrations for \lamr colors where $\lambda$ refers to \textit{ugizJHK}. 
We use $r$ as the basis for colors because the photometry in that band is 
the most precise of all bands. 
Note that although the focus of this paper is on the NIR extinction curve, we still need reddenings in optical bands to determine \Euz (our proxy for dust column density) and \Rv.
We require there to be at least 10 calibrator stars (\EBVcutlo) in a group for the regression to take place (see Section \ref{subsec:sample} for details on cuts). 
The intrinsic colors are fit using
\begin{align} \label{eqn:regression}
\begin{split}
    (\lambda-r)_0 &= a_0(\log g -4) + a_1(\log g -4)^2 \\
    & + a_2(\text{[Fe/H]}+1) + a_3(\text{[Fe/H]}+1)^2 \\
    & + a_4(T_\text{eff}-6000) \\
    & + a_5E(B-V)_\text{SFD} 
    + a_6.
\end{split}
\end{align}

Several other studies have used stellar parameters of large number of stars to infer intrinsic colors, and we highlight some of the differences in our approach.
\cite{Wang2019} used $\log T_\text{eff}$ and metallicity [M/H] (along with their second-order terms), as well as \logg. 
They, however, did not split by spectral class (their sample contained only clump giants).
For our sample where we split into spectral classes, a second-order term in \teff is not required.
Furthermore, \citet{Wang2019} 
consider as dust free the bluest 5\% of stars at a given \teff, whereas we use SFD dust maps.
Note also in our formulation the inclusion of an \sfdebv term, the purpose of which is to account for the fact that, strictly speaking, these calibrator stars have some (very small) reddening.
\citet{Schlafly2011} also use SSPP parameters as the basis for their intrinsic color prediction, but the correspondence between the parameters and colors comes from spectrophotometric models of stars, whereas in our case it is based on empirically selected dust-free stars.

We use the Huber Regressor as implemented in the Python package \texttt{scikit-learn} \citep{scikit-learn} to determine the coefficients in Equation \ref{eqn:regression}. This regression model employs the Huber loss function \citep{Huber1964}, which allows it to be less sensitive to outliers than more standard loss functions, while not ignoring them altogether. The weights are set to $1/\sigma^2$, where $\sigma$ is the photometric error on the color. We set the parameter $\epsilon=3.5$, which is similar to $3.5\sigma$ outlier rejection.

\begin{figure*}[t!]
    \centering
    \includegraphics[width=\linewidth,keepaspectratio]{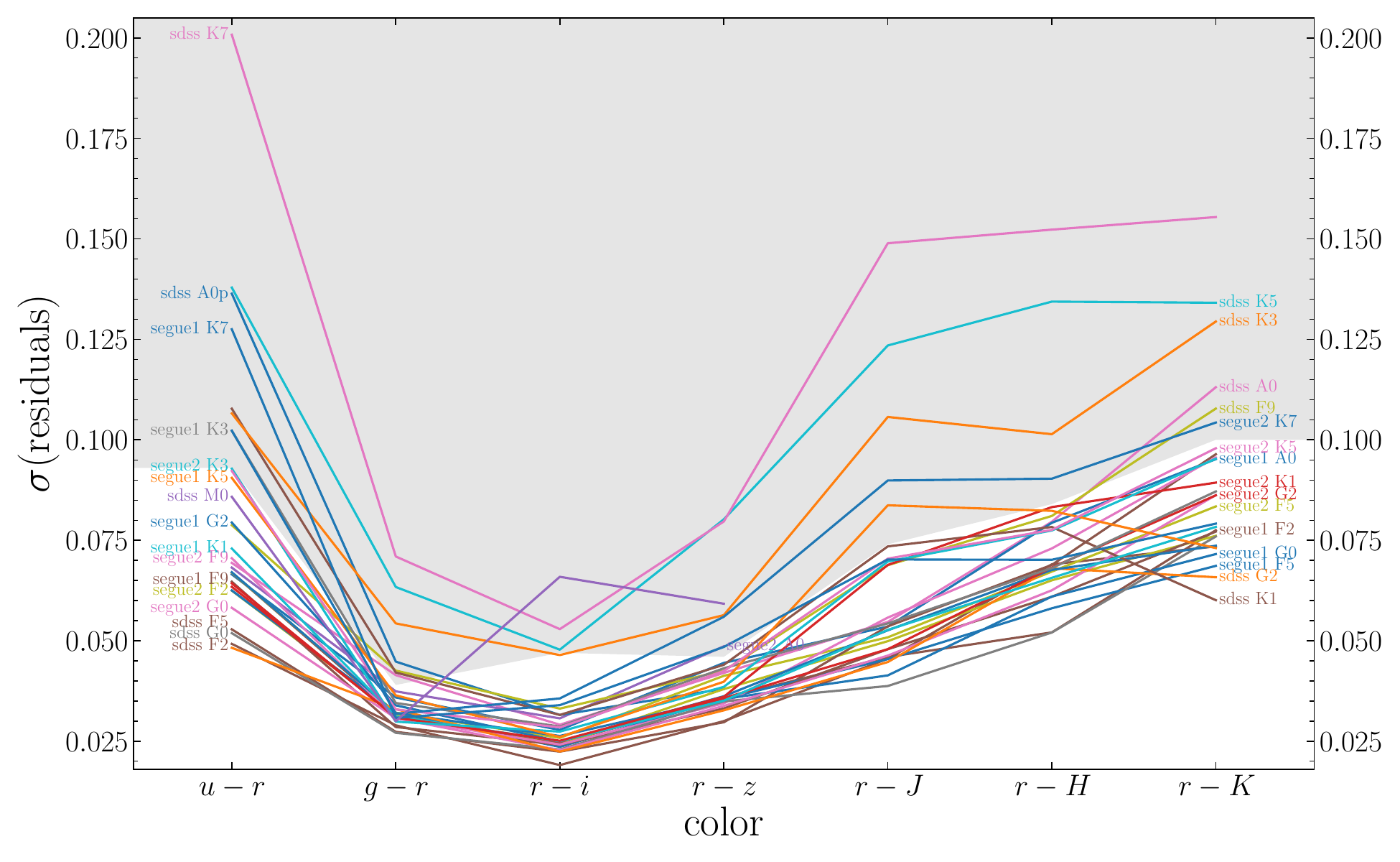}
    \caption{Standard deviation of the residuals vs. color for calibrator stars (those with \EBVcutlo). For each color, all survey/spectral subclass combinations in our original sample with at least 10 stars at \Euz are shown. The residuals come from regression to Equation \ref{eqn:regression}. 
    } 
    \label{fig:stdresid}
\end{figure*}

The regression produces one set of 7 coefficients for each color within each \survtype grouping. We calculate the standard deviation of the residuals (\stdresid) in each case as an assessment of the constraining power of that group or stars (which is some combination of fit quality and photometry errors). 
Figure \ref{fig:stdresid} gives an idea of how \stdresid varies across the range of colors and groups. Each line corresponds to a \survtype group. Residuals are the smallest in the optical bands, and in each color the majority of groups have similar residuals. 
In certain cases, the residuals are large enough that the corresponding group of stars warranted exclusion from the analysis. While there is no absolute sense in which \stdresid can be used to exclude a group, we used a relative assessment within each color. 
For a given color \lamr, we plot a cumulative distribution function (CDF) of the number of program stars (the ones to which the calibration will be applied) ranked by \stdresid. 
This allows us to eliminate groups which would provide only a modest increase in final program sample size, but at the cost of relatively large \stdresid. We make note of a cutoff \stdresid value for each \lamr, and exclude that \survtype group in that color from further analysis. 
The gray-shaded region in Figure \ref{fig:stdresid} indicates where we remove the grouping for a given color based on the CDF method described above. 

\begin{figure*}[t!]
    \centering
    \includegraphics[width=0.9\linewidth,keepaspectratio]{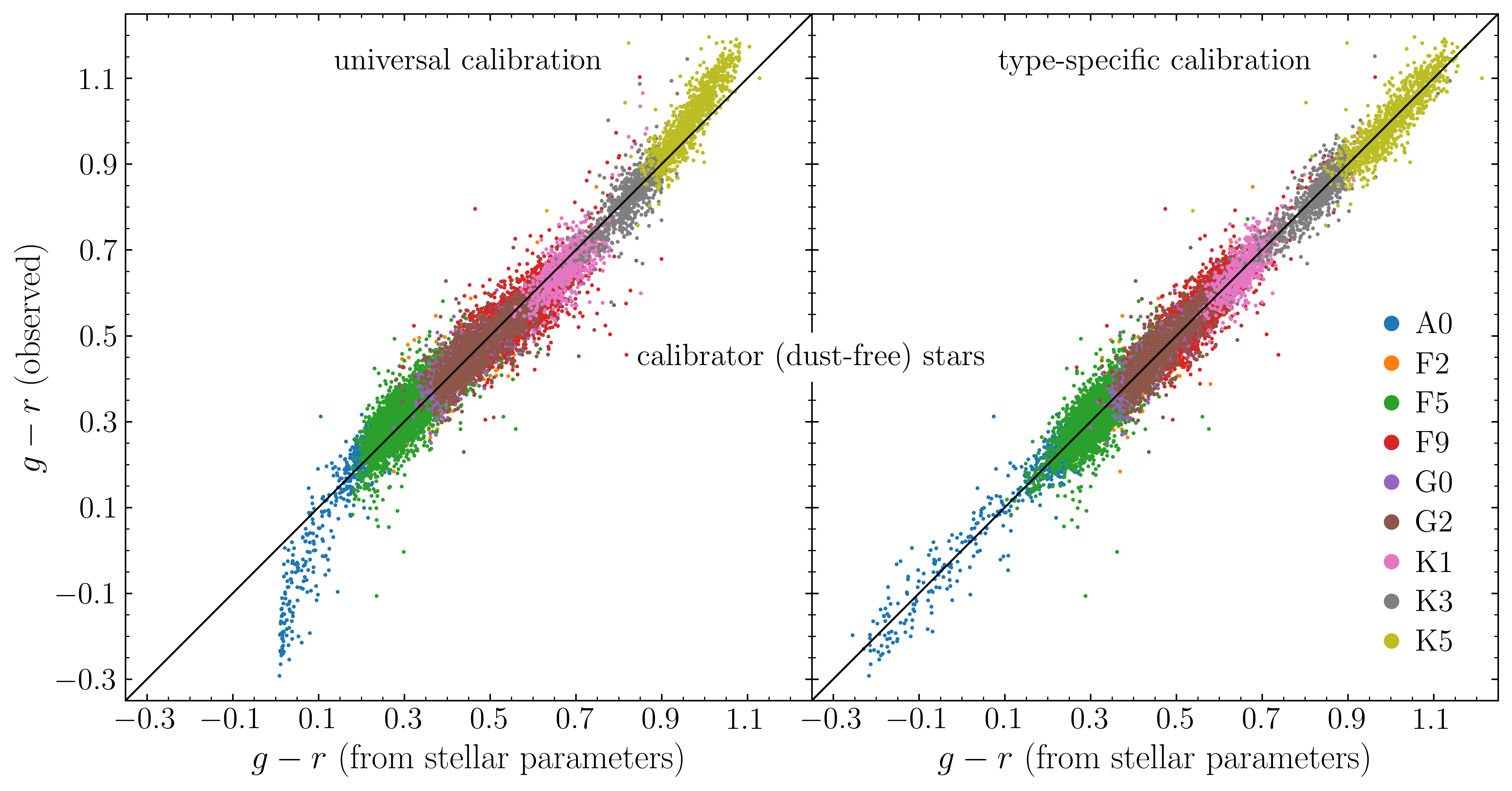}
    \caption{Observed SDSS $g-r$ color vs. $g-r$ color predicted from stellar parameters, using Equation \ref{eqn:regression} and the method described in Section \ref{subsec:regression}. 
    Shown are dust-free stars from which calibrations are produced, i.e., for which the observed color is essentially the intrinsic color.
    \textit{Left:} The case where calibrations are based solely on stellar parameters (not used in this work). We used a $\log T_\mathrm{eff}$ instead of \teff here, along with an added second-order term in \teff. See text near the end of Section \ref{subsec:regression}. \textit{Right:} The calibrations used in this work, derived from stellar parameters \textit{and} done individually for each \survtype group.
    } 
    \label{fig:g-robspred}
\end{figure*}

In Figure \ref{fig:g-robspred} we show the observed $g-r$ colors vs. $g-r$ colors predicted using the stellar-parameter-based regression method. Only the calibrator stars are shown. 
The left panel shows what the relation would look like if the calibration was applied without separating by stellar subclass (universal calibration). 
We modified the model by replacing \teff with $\log T_\mathrm{eff}$, and added a second-order term in that parameter. These changes (especially the addition of the second-order term) improved the universal calibration beyond its performance using Equation \ref{eqn:regression} as-is.
However, there is still some curvature to the locus, with both A0 and K5 stars having appreciably poor predictions. There is also a larger spread in the middle of the locus than in the right panel, which shows our adopted type-specific calibration.
Aside from a very small number of outliers, our method appears to work remarkably well across a range of observed $g-r$. 
As mentioned, we also split by "survey" (one of three spectroscopic campaigns under the umbrella of SDSS: the Legacy component, which we simply refer to as SDSS, along with SEGUE1 and SEGUE2). We find that given a fixed subclass, there is still subtle variation in the calibration among these; however, the effect is much smaller than the one from the subclasses themselves illustrated in Figure \ref{fig:g-robspred}.


\begin{figure*}[p!]
    \centering
    \includegraphics[width=\textwidth,height=\textheight,keepaspectratio]{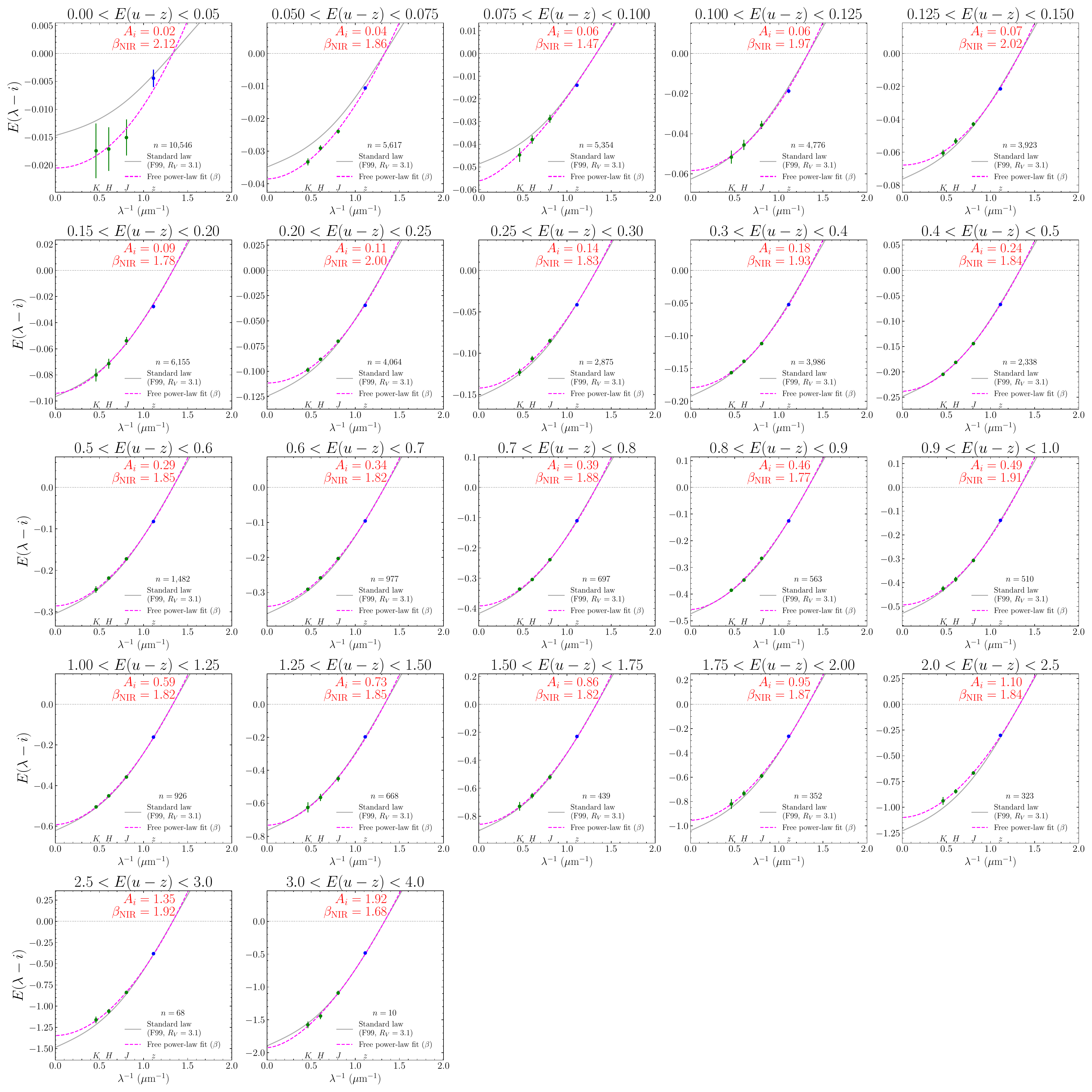}
    \caption{\Elami vs. \invlam in 22 bins of \Euz. The magenta dashed curves represent best fits to the two-parameter model described by Equation \ref{eqn:2param}. The gray curves represent a standard \cite{Fitzpatrick1999} extinction law with \Rvstd based on the mean \Euz of the bins. Error bars on the points are rescaled such that $\chi^2_\text{red}\equiv 1$. The central wavelengths for each of the four filters incorporated in the fits are labeled along the $x$-axis. Fitted \bnir and \ai values are shown near the top of each panel, above the $E(\lambda-i)=0$ dotted line. 
    }
    \label{fig:NIR22panel}
\end{figure*}

\subsection{Color Excesses (Reddenings)} \label{subsec:colex}

For each program star, 
i.e., those with \EBVcuthi in \survtype groupings which were not excluded by the CDF cutoff (see Section \ref{subsec:regression})
we use its stellar parameters with Equation \ref{eqn:regression} to determine intrinsic color (\lamr). We set $E(B-V)=0$, since we want the predicted color of the star if it had exactly zero dust along it sightline. We then subtract these predicted intrinsic \lamr from the observed colors to produce a set of color excesses: $E(\lambda-r)=(\lambda-r)-(\lambda-r)_0$. 

We also calculate \Euz as 
\begin{equation*}
    E(u-z) = E(u-r) + E(r-z),
\end{equation*}
which we consider to be a proxy for the dust column. 
If extinction curves were identical across the sky (i.e., $R_V=\text{const.}$), any color excess would be equally well correlated with the dust column, and one would pick whichever reddening is best measured, or is closest to the traditional \ebv. However, the extinction curve is not universal; \citet{Butler2021} showed that $A_u$ correlates with \nh, and therefore the dust column, better than any other optical/NIR extinction. This, combined with the fact that $z$ is the optical band least affected by dust, leads us to choose \Euz as the best available proxy for the dust column.

\subsection{NIR Extinction Curves} \label{subsec:extcurves}

There are two principal ways in which we construct curves: (1) for individual stars that have higher reddening (low Galactic latitude regime), and (2) for aggregated stars. In the low-dust/high-latitude regime, many individual stars have very uncertain color excesses, making curve derivation meaningless. For that reason, in method (2), we instead use the power of our large sample to aggregate the stars by averaging their color excesses in bins of dust along their sightline (namely, its proxy \Euz). 

To determine the parameterized NIR extinction curve, we choose a power law, which is known to provide a good description of the extinction curve between 0.75 and $\sim$5 \um \citep[e.g.,][]{Martin1990, Fitzpatrick2009, Wang2019}. In our dataset this range includes bands $i$ through $K$. 
We switch our basis (or "anchor") from $r$ to $i$: $E(\lambda-i)=E(\lambda-r)-E(i-r)$, 
because $i$, unlike $r$, belongs to this power-law region of the extinction curve.
The power law formulation of the reddening curve is
\begin{equation} \label{eqn:2param}
    E(\lambda-i) = A_i \left[\left(\frac{\lambda}{\lambda_i}\right)^{-\beta_\text{NIR}}  -1 \right],
\end{equation}
where $\lambda_i=7510$ \AA.
The two free parameters are the slope of the power law (\bnir) and the extinction in $i$ band (\ai), the latter of which allows for any reddening \Elami to be transformed into an absolute extinction \alam. 
Technically, the extinctions in Equation \ref{eqn:2param} should be monochromatic, whereas we use bandpass extinctions. However,
in the optical and NIR, extinctions within bandpasses are close to monochromatic extinctions at their central wavelengths. 

To generate curves from aggregated stars, we create 22 bins in \Euz, the edges of which are given in the titles of the panels in Figure \ref{fig:NIR22panel}. The bins become progressively wider with increasing \Euz due to the majority of stars being at low dust. 
For 18\% of the stars in our sample, the \Euz value we determine is negative, 
which can be attributed to the inherent uncertainties in the determination of intrinsic colors (and therefore color excesses) of stars which have very small amounts of dust along their sightlines. Such stars are not included in the analysis.

In each \Euz bin, we fit Equation \ref{eqn:2param} to the median of the \Elami values of the individual stars. 
We restrict our sample in a given bin to only those stars which have a valid magnitude and error measurement in all four bands \textit{zJHK} (in addition to passing the cuts in Section \ref{subsec:sample}). 

The fits are carried out using the unweighted non-linear least squares method, as implemented in the \texttt{curve\_fit} function within the \texttt{scipy} \citep{scipy} package in Python. We return $A_i$ and $\beta$ values along with their errors. 
We choose to perform the fits without weights because the proportionally much smaller error on $E(\lambda-z)$ disproportionately affects the fits (as compared to its NIR counterparts).
We determine the effective errors of \Elami asking that they agree with the power-law model. 
To do so, we calculate a \redchisq value for each fit, and rescale the nominal errors of the median \Elami such that they produce $\chi^2_\text{red}\equiv 1$. The nominal error of the median \Elami is determined as the standard deviation of individual \Elami values divided by $\sqrt{n}$, where $n$ is the total number of stars in the bin.

To derive parameterized extinction curves for individual stars, we use much the same method as described above. 
We also derive $R_V$ values for both aggregated and individual stars. 
To summarize, we derive NIR extinction curves parameterized by the power-law slope \bnir (with normalization \ai), as well as \Rv, for stars aggregated in 22 reddening bins spanning $0<E(u-z)<4$, and for 2700 individual stars with \Euzcut.
We use the \citet{Fitzpatrick1999} extinction law at \Rvstd to convert between reddenings in SDSS colors to reddenings in Johnson colors, finding $E(B-V)=0.86E(g-r)$ and $E(V-r)=0.46E(g-r)$. To obtain $R_V=A_V/E(B-V)$, we substitute $A_V=A_i+E(r-i)+E(V-r)$.
This conversion is somewhat sensitive to the choice of extinction law. We use the \citet{Fitzpatrick1999} extinction law rather than e.g. \ctccm because the former has been shown to correctly predict $E(g-r)$ \citep{Schlafly2010}, whereas the \ctccm prediction is about 20\% too small.


\section{Results} \label{sec:results}

\subsection{NIR Extinction Curve vs. the Dust Column}
\label{subsec:bnirEuz}

For diffuse dust, extinction curves should not depend on the dust column alone, unless the dust composition and grain size distribution are different (which in principle they can be between high and low latitudes). 
However, this expectation has not been tested in the NIR at high and low latitudes simultaneously.
The panels in Figure \ref{fig:NIR22panel} show aggregate NIR reddening as a function of inverse wavelength, where each panel is a different bin of the dust column, \Euz. The dust column increases to the right and downward within the set of panels. In each panel we show the best fitting power-law curves and their parameters. Also shown for reference are standard \cite{Fitzpatrick1999} extinction curves with \Rvstd, scaled to match the mean \Euz of the bin. 
The \cite{Fitzpatrick1999} curve is not defined as a strict power law in any wavelength regime. In some bins it matches the observations better than in others, without any particular pattern.
Note that the range of \Elami increases with each succeeding bin. We can see that the power law represents a good description of reddening regardless of dust column density.

\begin{figure}[t!]
    \centering
    \includegraphics[width=\linewidth,keepaspectratio]{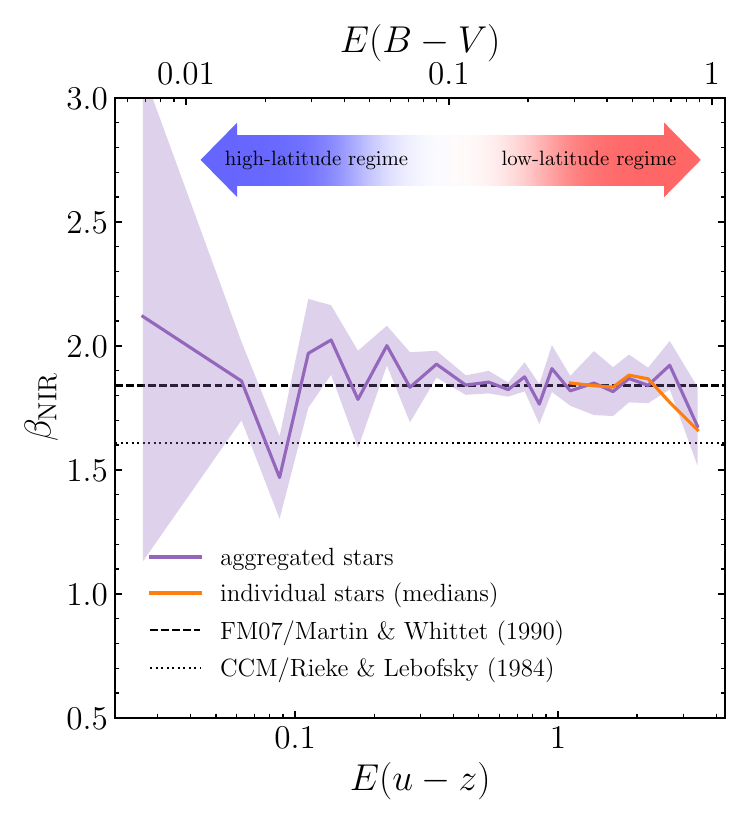}
    \caption{Power-law slope of the NIR extinction curve (\bnir) vs. a proxy for the dust column (\Euz) for the fits described in Section \ref{subsec:extcurves}. The purple line and associated shaded error region shows the values determined from the 22-bin aggregated fits as shown in Figure \ref{fig:NIR22panel}. The orange line shows individual stars, which were then binned \textit{after} fitting for \bnir and \ai. Two horizontal dotted lines are included, showing \bnir values from impactful studies: 1.84, from \cite{fm07} \citep[originally from][]{Martin1990}, and 1.61, from \cite{CCM} \citep[originally from][]{Rieke1985}. Note that \citet{Fitzpatrick1999} find $\betamath\approx1.62$. The upper secondary $x$-axis is $E(B-V)=0.26 E(u-z)$. The arrows indicate an approximate separation between the high- and low-latitude regimes in our study, to either side of $E(B-V)\sim 0.1$. See Figure \ref{fig:lat}.  
    } 
    \label{fig:betaNIR}
\end{figure}

In this study we are particularly interested in how \bnir may vary with the dust column, which is shown in Figure \ref{fig:betaNIR}. The purple line represents the \bnir values determined by the fits to aggregated (binned) stars, as described above. We also include a secondary $x$-axis with \ebv, for convenience and ease of comparison with other studies. This \ebv is related to \Euz by the simple scaling $E(B-V)=0.26 E(u-z)$, determined from all stars in our sample. Note that the reddening is presented on a log scale. A large portion of the horizontal span is occupied by rather minuscule reddening.

It is clear from Figure \ref{fig:betaNIR} that there is no strong correlation between \bnir and the amount of dust along a sightline. As a further test, we show, in the same bins, the median power-law slopes of individual stars with $E(u-z)>1$ (see Section \ref{subsec:extcurves}). 
Medians of \bnir based on extinction curves for individual stars (orange line in Figure \ref{fig:betaNIR}) agree with those from the aggregated stars.

Since we find no correlation between \bnir and the dust column, we find it appropriate to report a single value for our study: $\beta_\text{NIR} = 1.85 \pm 0.01$. This value is determined by taking a weighted average of the \bnir values from the 22 bins. We report it along with separate values for high- and low-latitude regimes (corresponding to low and high values of \Euz, respectively) in Table \ref{tab:beta}, which are again consistent with a single overall value. 

\begin{figure*}[t!]
    \centering
    \includegraphics[width=\textwidth,keepaspectratio]{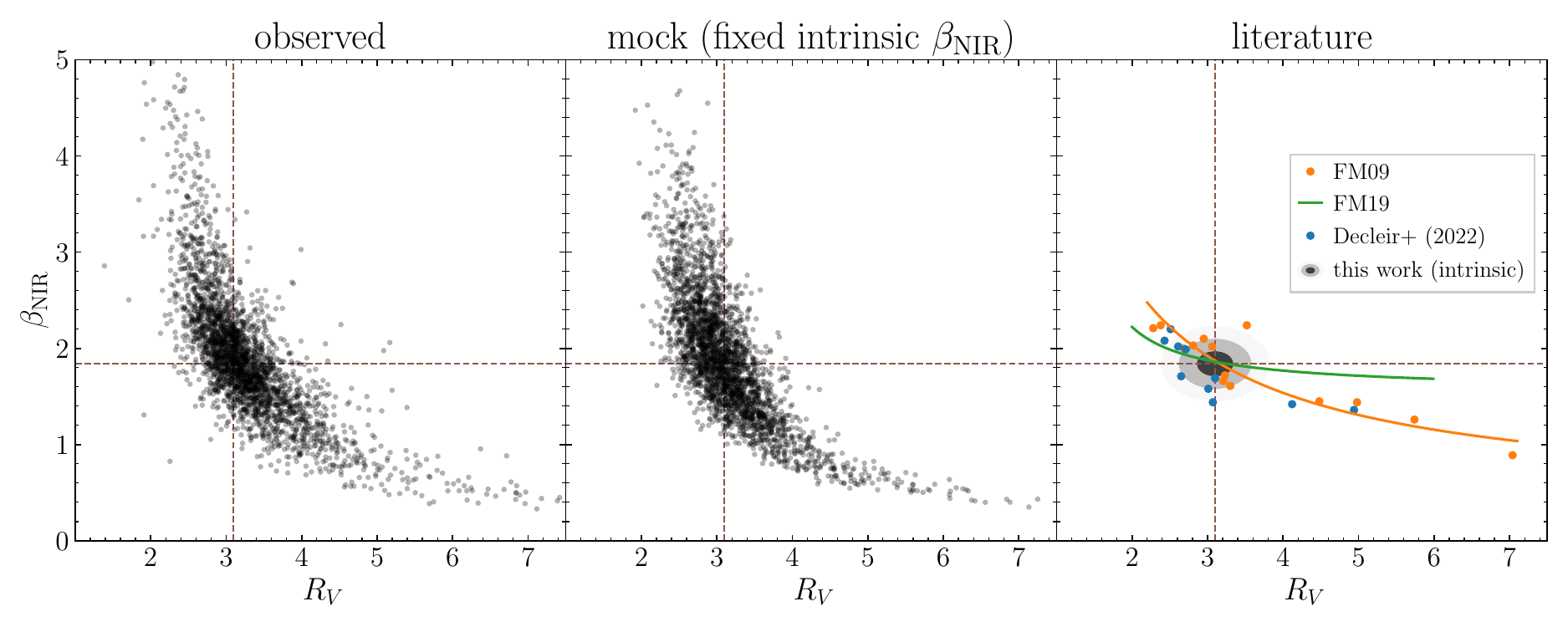}
    \caption{Power-law slope of the NIR extinction curve (\bnir) vs. a representation of the optical slope of the extinction curve (\Rv) for three cases. In each case, standard literature values of \Rvstd \citep[e.g.,][]{CCM} and \bnirstd \citep{Martin1990, fm07} are included as brown dashed lines. 
    \textit{Left:} Observed individual stars with $E(u-z)>1$. 
    \textit{Center:} Analogous to the left panel, but using a comparably sized sample of mock stars generated from the \cite{Fitzpatrick1999} family of curves, which are consistent with a fixed \bnir. See Section \ref{subsec:bnirRv} for details. 
    \textit{Right:} Selected literature values for sightlines with measured \bnir and \Rv. The orange points and associated fit come from \cite{Fitzpatrick2009}. 
    The purple points come from \citet{Decleir2022}. 
    The gray shaded ellipses represent $1\sigma$, $2\sigma$, and $3\sigma$ contours of a two-dimensional Gaussian with $\sigma(R_V)=0.24$ and $\sigma(\betamath)=0.13$. These are intrinsic values derived from mock comparisons to observed data in the tails of the distribution (see Section \ref{subsec:bnirRv}).
    }
    \label{fig:betaNIR_Rv}
\end{figure*}

\subsection{NIR Extinction Curve vs. {\normalfont \Rv}}
\label{subsec:bnirRv}

UV and optical extinction curves are known to vary as a function of \Rv, but whether that extends to the NIR regime has not been firmly established.
The relation we find between \bnir and \Rv is plotted in the leftmost panel of Figure \ref{fig:betaNIR_Rv}. 
Note that in this analysis we use individual (high reddening) extinction curves, since little variation in \Rv (and \bnir) is seen for aggregated stars.
There is a strong apparent correlation between \bnir and \Rv. At the average Milky Way value of \Rvstd, the points reassuringly passes quite near the commonly used \bnirstd \citep{Martin1990, fm07}, but the ranges in both \bnir and \Rv are quite large.
Although the correlation between \bnir and \Rv appears quite strong based on formal values, we will show that it is not a true reflection of the intrinsic relationship between the two parameters. 

The \bnir and \Rv values we determine for individual sightlines have associated uncertainties which stem from photometric and spectroscopic observations as well as our intrinsic color estimation process (Section \ref{eqn:regression}). 
\citet{Larson2005} find that, for low-extinction sightlines, relatively small photometric errors can produce very large errors in \Rv. 
Furthermore, although \bnir and \Rv are independent quantities, the former being the slope of the extinction curve in the NIR whereas the latter being related to the slope in the optical ($1/R_V = A_B/A_V -1$), they both depend on the determination of the absolute extinction, which via Equation \ref{eqn:2param} is correlated with \bnir. 
Indeed, \citet{Fitzpatrick2009} report error bars on the sightlines in their \bnir vs. \Rv relation (see their Figure 4) which are not independent, and follow the general shape of the correlation we see in Figure \ref{fig:betaNIR_Rv}. To test the extent to which the apparent correlation is driven by correlated errors, we perform an analysis using mock observations.

The idea of the mock analysis is the following. We will assume that the actual extinction curves have identical \bnir irrespective of \Rv. We will then add realistic noise to the reddenings that these fixed-beta model curves predict and see what values of \bnir (and \Rv) we recover. It is therefore important to distinguish between true (or input) \bnir and \Rv and the ones we derive ("observe") due to the uncertainties in reddening.

The mock analysis uses the same sample size as we have for individual stars. We generate curves by assuming the \citet{Fitzpatrick1999} law, which essentially has an invariant NIR extinction curve, and therefore no dependence on \Rv. The input (intrinsic) \Rv values for the mock curves are drawn from a Gaussian distribution with mean \Rvstd. The width of the distribution of intrinsic \Rv values is much smaller than the observed range and will be discussed shortly.

\begin{deluxetable}{cccc}

    \tablecaption{Values for the slope of the NIR extinction curve. 
    \label{tab:beta}}
    
    \tablehead{\colhead{Regime} & \colhead{$\beta_\text{NIR}$} & \colhead{$n_\text{stars}$} & \colhead{$E(B-V)$}} 
    
    \startdata
    High latitude & $1.91\pm0.04$ & \hilatstars & 0 -- 0.1 \\
    Low latitude & $1.84\pm0.02$ & \lolatstars & 0.1 -- 1 \\
    \tableline
    All stars & $1.85\pm0.01$ & \allstars & 0 -- 1 \\
    \enddata

\end{deluxetable}

For each real star, we thus have its mock \Elami values for $\lambda=(z,J,H,K)$. We perturb these color excesses by an amount derived from the uncertainties of the real stars. 
We then employ the same method as was used in Section \ref{subsec:extcurves} to fit the power-law extinction model (Equation \ref{eqn:2param}) and obtain "observed" \bnir and \ai values. The output \Rv values of the mock stars are derived in the same way as in the observed case. 
A slight "zero-point" adjustment is made to the derived mock \bnir in order to take into account that \cite{Fitzpatrick1999} assumes a shallower beta than our overall value and is not a perfect power-law over the \textit{izJHK} range.

As can be seen in the middle panel of Figure \ref{fig:betaNIR_Rv}, the shape of the mock distribution bears striking resemblance to the shape of the observed distribution (left panel). This similarity strongly indicates that the apparent correlation we find between \bnir and \Rv for observed stars is an artifact of correlated uncertainties, and not the nature of the \textit{intrinsic} relationship between the parameters.
In order for the mock to most closely match the observed distribution, we have allowed the input \Rv to have a range of values drawn from a Gaussian distribution with a standard deviation of 0.28.
Note that this intrinsic spread in \Rv is much smaller than the spread of mock "observed" \Rv values (the ones that result from taking observational uncertainties into account). 

In reality, both the intrinsic \Rv and intrinsic \bnir can vary. 
Armed with the information that the intrinsic span in \Rv is small, we refine the value of intrinsic \sigRv and determine \sigbeta in the following way. We add various amounts of scatter in both \Rv and \bnir to the mock distribution after it has been generated using a fixed \Rvstd. When the scatter in the output \Rv values in the upper tail of the mock distribution (where the horizontal scatter is dominated by \Rv) equals the scatter in the same part of the observed distribution, and the same is true for the scatter in \bnir in the lower tail (where the vertical scatter is dominated by \bnir), we take those scatters to be the intrinsic. In this way, we get $\sigma(R_V)=0.24$ and $\sigma(\betamath)=0.13$. We schematically plot these derived intrinsic distributions using 2D Gaussian contours in the right panel of Figure \ref{fig:betaNIR_Rv}. We tested a scenario where the intrinsic \bnir and \Rv were linearly correlated, but found that values with correlation coefficients other than 0 were taking us away from the observed distribution. We therefore conclude that intrinsically \bnir and \Rv not only have a much smaller range than what the nominal values suggested, but also seem to be uncorrelated.

We also tested the correlation found in \citet{Fitzpatrick2009} as opposed to a simple linear correlation. The run which matched the observed distribution the best had a similar spread in \Rv, but deviated $\sim$10\% more in \bnir spread than the best match with no intrinsic correlation assumed. Our sightlines, with a relatively small range of $R_V$ values, thus disfavor a \citet{Fitzpatrick2009} (or any other) correlation.

\begin{figure}[t!]
    \centering
    \includegraphics[width=\linewidth,keepaspectratio]{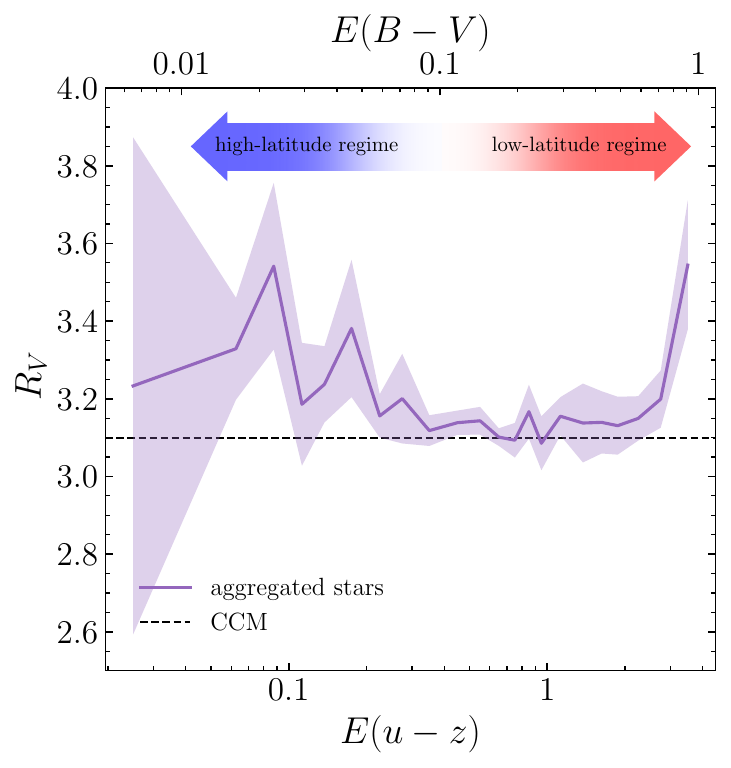}
    \caption{$R_V \equiv A_V/E(B-V)$ vs. a proxy for the dust column (\Euz) using the same bins and fits as Figure \ref{fig:betaNIR}. The dotted line indicates the accepted MW value \Rvstd \citep[e.g.,][]{CCM}. The upper secondary $x$-axis is $E(B-V)=0.26 E(u-z)$. The arrows indicate an approximate separation between the high- and low-latitude regimes in our study, to either side of $E(B-V)\sim 0.1$. See Figure \ref{fig:lat}.
    }
    \label{fig:RvEuz}
\end{figure}

\subsection{{\normalfont \Rv} vs. the Dust Column} \label{subsec:RvEuz}

In addition to testing for a correlation between \Rv and \bnir, we also investigated whether \Rv might be dependent on the dust column density along sightlines, represented here by \Euz. The resultant plot, Figure \ref{fig:RvEuz}, is analogous to Figure \ref{fig:betaNIR}, simply replacing \bnir on the $y$ axis with \Rv. We again use the aggregated stars with the same bins as in Figure \ref{fig:NIR22panel}. The \Rv values are determined using the same conversion as described at the end of Section \ref{sec:methods}, this time employing the median $E(g-r)$ in each bin (alongside median \Eri and fitted $A_i$), resulting in a median-based \Rv.

One may in principle not expect \Rv to vary with the dust column, since \Rv is generally thought to track the size distribution of grains. However, in denser regions, the size distribution may change (toward larger grains) due to higher probability of accretion and coagulation \citep[e.g.,][]{Draine2003}. 
It is important to note that density, when applied to the discussion of size distribution, is physical as opposed to a column density, since physical density is what drives grain interaction. Column density may happen to track physical density, but for a given column density, one can imagine several different dust distribution scenarios along a line of sight (e.g., clouds interspersed with relative voids versus a constant diffuse distribution).

Since we do not probe the densest regions in this work (at least in terms of column density), we may not expect to see any rise in \Rv with the dust column. At the lower dust columns associated with high Galactic latitudes, the expectation is less clear. \citet{Schlafly2016} produced a relationship between their $R_V'$ and $E'$ (roughly equivalent to \Rv and \ebv) in their Figure 16. We can compare the trends in that figure directly with those in our Figure \ref{fig:RvEuz}. At the highest \ebv, they find a rather sharp rise in \Rv (as may be expected per the above discussion), and we do as well in our highest-\Euz bin. However, the maximum \ebv we probe is significantly lower than theirs ($\sim$1 vs. $\sim$2.5). There is evidence for a slight rise around $E(B-V)=1$ in their results, but not of nearly the magnitude we find. Since our densest bin has only 18 stars, we take the sharp rise there with a grain of salt (although there may be some expectation for a change around $E(B-V)=1$, as \citealp{Whittet1988b} find the evidence that ice mantles form in almost all cases above that point).
\citet{Schlafly2016} note a large increase in uncertainty below $E(B-V)=0.5$, but seem to show a rise below roughly that point; we find increased \Rv below $E(B-V)\sim 0.1$, albeit with large uncertainties at the very lowest dust columns.

Overall, given the low number of stars in the highest-dust bin and the relative unreliability of the results in the lowest 2-3 bins, we seem to find rough agreement with the established MW standard \Rvstd across a range of dust columns. There is some evidence for a rise toward high latitudes, but it is fairly weak.

\begin{figure}[t!]
    \centering
    \includegraphics[width=\linewidth,keepaspectratio]{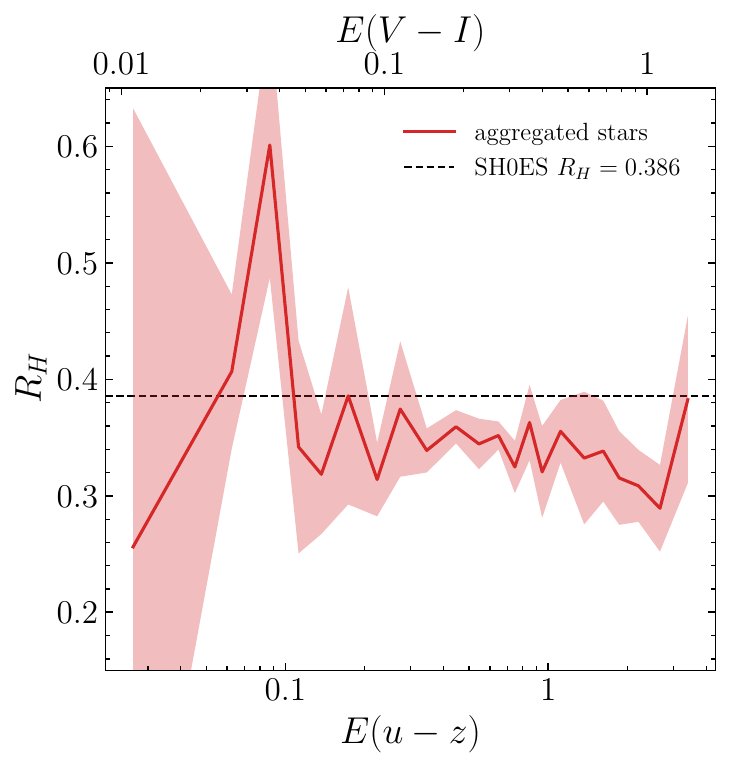}
    \caption{$R_H\equiv A_H/(A_V-A_I)$ vs. a proxy for the dust column (\Euz) using the same bins and fits as Figure \ref{fig:betaNIR}. The dotted line indicates the adopted value of \RHstd used by the SH0ES team in Cepheid dereddening en route to the derivation of \hub \citep[e.g.,][]{Riess2016, Riess2022}. The upper secondary $x$-axis is $E(V-I)=0.42 E(u-z)$.
    }
    \label{fig:RHEuz}
\end{figure}

\begin{figure*}[t!]
    \centering
    \includegraphics[width=\textwidth,keepaspectratio]{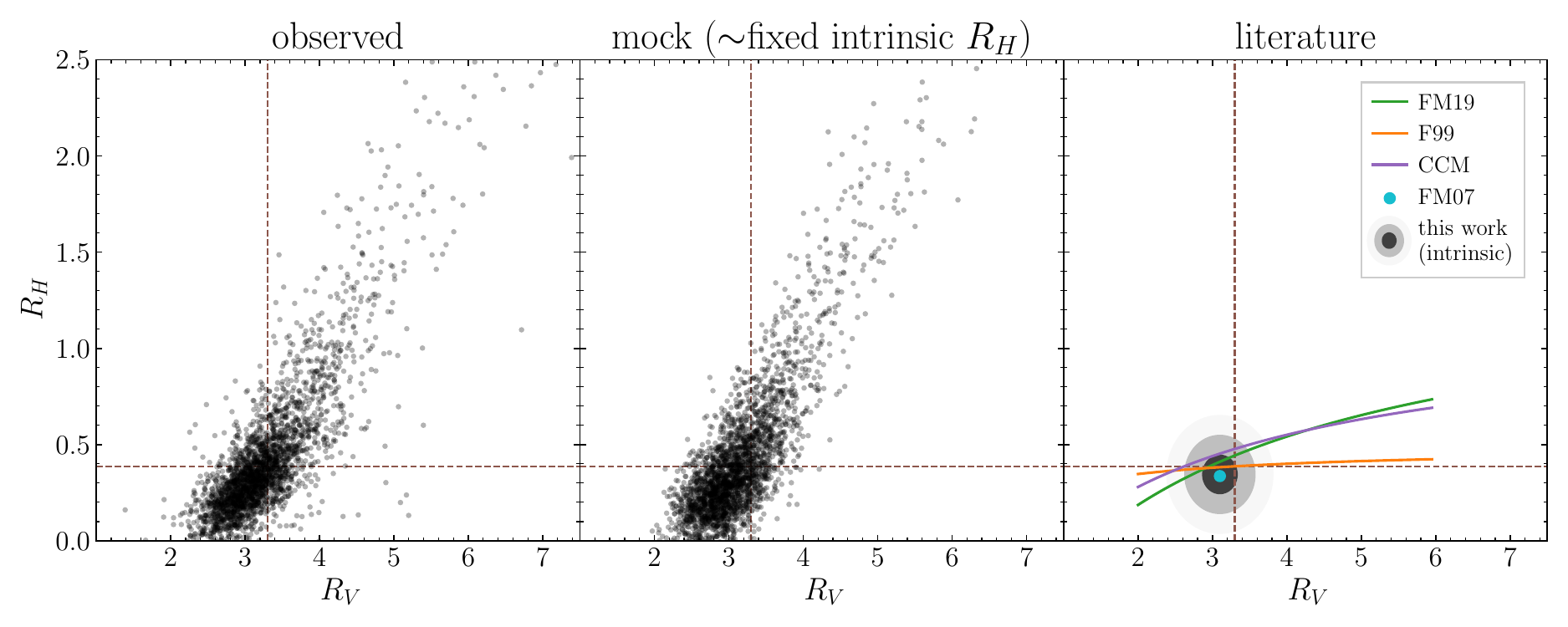}
    \caption{$R_H\equiv A_H/(A_V-A_I)$ vs. a representation of the optical slope of the extinction curve (\Rv) for three cases. In each case, standard literature values of $R_V=3.3$ and \RHstd \citep{Riess2016,Riess2022} are included as brown dashed lines. These panels are directly analogous to the three \bnir panels, but show \RH instead.
    \textit{Left:} Observed individual stars with \Euzcut. 
    \textit{Center:} Analogous to the left panel, but using a comparably sized sample of mock stars generated from the \cite{Fitzpatrick1999} family of curves, which are consistent with a fixed \bnir (and roughly fixed \RH). See Section \ref{subsec:bnirRv} for details.
    \textit{Right:} Values calculated from the \citet[][FM19, green]{Fitzpatrick2019}, \citet[][F99, orange]{Fitzpatrick1999}, \ctccm (purple), and \citet[][FM07, cyan, \Rvstd]{fm07} literature extinction laws. The gray shaded ellipses represent $1\sigma$, $2\sigma$, and $3\sigma$ contours of a two-dimensional Gaussian with $\sigma(R_V)=0.24$ and $\sigma(R_H)=0.10$. These are intrinsic values derived from mock comparisons to observed data (see Section \ref{subsec:H0}).
    } 
    \label{fig:RHRv}
\end{figure*}

\subsection{The NIR Extinction Curve and Cepheid Calibration} \label{subsec:H0}

Recent Cepheid period-luminosity calibrations have been performed in the near-IR, in particular the $H$ band. Whether the calibration is performed using Wesenheit magnitudes \citep{Madore1982}:
\begin{equation}\label{eqn:mhw}
    m_H^W \equiv m_H - R_H (V-I)
\end{equation}
or unreddened magnitudes:
\begin{equation}\label{eqn:mh0}
    m_H^0 \equiv m_H - R_H E(V-I),
\end{equation}
one needs to know \RH, an extinction curve-dependent parameter, precisely.

Since modern measurements of the Hubble constant are based on Hubble Space Telescope photometry,  $H$, $V$, and $I$ here refer to the Hubble filters \Hhub, \Vhub, and \Ihub, respectively. The parameter $R_H$ (elsewhere called simply $R$ or sometimes $R_\text{E}$) is of similar makeup to the more familiar \Rv:
\begin{equation} \label{eqn:RH}
    R_H = \frac{A_H}{E(V-I)}
\end{equation}
The most recent studies from SH0ES \citep{Riess2016,Riess2022} use for their baseline fit the value \RHstd, derived from the \cite{Fitzpatrick1999} reddening law assuming a somewhat higher than standard \Rvhub. 
Given that law, $R_H$ only changes by $\sim$0.005 between \Rvhub and \Rvstd, so the choice of \Rv is not of great importance. 
The use of a constant \RH for any Cepheid is equivalent to assuming a universal power-law slope in the NIR. Therefore, based on the analyses in Sections \ref{subsec:bnirEuz} and \ref{subsec:bnirRv}, we would expect \RH to be constant as well, but we still need to verify that and determine its value.

With a straightforward reparameterization of our results, we can investigate how \RH may vary with sightline parameters. To determine \RH from our framework, we again use \citet{Fitzpatrick1999} to construct transformations between reddenings, here converting between SDSS bands and HST bands. We find 
\begin{equation} \label{eqn:EVI}
    E(V-I) = 0.31E(g-i) + 1.43E(r-i)
\end{equation}
We obtain \AH to be used in Equation \ref{eqn:RH} by adjusting the \AH as defined in 2MASS photometry as
\begin{equation}
    A_H = A_{H, \text{2MASS}} \left(\frac{\lambda_{H,\text{F160W}}}{\lambda_{H, \text{2MASS}}}\right)^{-\betamath},
\end{equation}
where we use $\betamath=1.85$ as determined in section \ref{subsec:bnirEuz}. 

In Figure \ref{fig:RHEuz}, we plot the relation between \RH and \Euz. With the exception of some large spikes at very low dust where we do not constrain it well, $R_H$ is essentially invariant with the dust column. We find the mean \RHuserr across the last 10 (low-latitude) bins, whose $E(V-I)$ range corresponds to that of the MW Cepheid sample presented in \citet{Riess2021, Riess2022}. 

We can also use the same mock analysis framework from Section \ref{subsec:bnirRv} to investigate potential variation of \RH with \Rv. The left panel of Figure \ref{fig:RHRv} indicates that, for individual sightlines ($E(B-V)>0.25$), there is quite a steep correlation between \RH and \Rv. However, we must again test whether the correlation is intrinsic by comparing to a mock dataset. The set of mock stars we use here is exactly the same as is shown in the middle panel of Figure \ref{fig:betaNIR_Rv}. This time, we simply calculate \RH using the method described above and plot those values vs. \Rv instead. Indeed, we see a similar correlation in steepness and direction as in the observed case. The observed trend of \RH with \Rv again appears to be a result of uncertainties in measurement. 

To establish an intrinsic scatter in \RH, we propagate the intrinsic scatter in \bnir (Section \ref{subsec:bnirRv}) and obtain $\sigma(R_H)=0.10$, depicted in the vertical direction of the 2D Gaussian shown in the right panel of Figure \ref{fig:RHRv}. 

\subsection{Recipe for Use} \label{subsec:recipe}

We recommend the following simple recipe by which to make use of our results in correcting NIR observations:
\begin{enumerate}
    \item Produce an extinction curve (\alam) based on some reddening \citep[e.g., \ebv from \ctsfd or][]{Schlafly2011} using a standard curve of choice. This curve can be \Rv-dependent, as in \ctccm, \citet{Fitzpatrick1999}, and \citet{Fitzpatrick2019}, or the fixed \Rvstd curve of \citet{fm07}. The Python packages \texttt{extinction} \citep{extinction} and \texttt{dust\_extinction} \citep{dust-extinction} may be useful for this, or \texttt{CCM\_UNRED} and \texttt{F99\_UNRED} in IDL. Note that \texttt{CCM\_UNRED} may actually use \citet{ODonnell1994} parameters as default.
    \item Beyond $\lambda=7500$ \AA, replace the standard curve \alam with
    \begin{equation}
        A_\lambda = A_{7500}\left(\frac{\lambda}{7500 \text{ \AA}}\right)^{-\betamath},
    \end{equation}
    where the relevant values of \bnir are given in Table \ref{tab:beta}.
\end{enumerate}

\section{Discussion} \label{sec:discussion}

\subsection{Slope of the NIR Extinction Curve} \label{subsec:disc_PL}

\subsubsection{Standard Extinction Laws} \label{subsubsec:standard}
There is a large number of studies, especially in recent years, that have investigated the NIR extinction law. Nonetheless, the NIR extinction curves that are most commonly used still come from the general UV through NIR extinction curves, especially the ones which have gained traction over the years and are referred to as the "standard" Milky Way curves. We start by comparing our 
average value for the power-law slope of the NIR extinction curve, $\betamath=1.85 \pm 0.01$, to other NIR power-law slopes associated with these standard curves.

The earliest standard curve still widely used is from \ctccm, which assumes a shallower curve than the one we find: $\betamath=1.61$, originally found by \citet{Rieke1985}. \citet{Rieke1985} investigated the extinction law in bands $J$ through $N$ (1 to 10 \um) toward a handful of M supergiants in the Galactic Center (GC), as well as toward $o$ Sco (an A5 II star at $b=18^\circ$) and Cyg OB2 \#12 (a B supergiant within an OB association in the Galactic plane). They found no variance in the NIR law between these sightlines.
So even though \ctccm curves are \Rv-dependent in the UV and optical, they are universal beyond 0.9 \um. 

The parameterization of another widely used law, from \citet{Fitzpatrick1999}, does not implement a strict power law in the NIR regime; rather, it makes use of fixed cubic spline anchor points at 1.2 and 2.7 \um, which
we find to also be close to $\betamath=1.6$ and thus shallower than our typical slope.
The sightlines on which the law is based are described in \citet{Fitzpatrick1990}. Their program sample consists of 45 OB stars within $\sim$20$^\circ$ of the Galactic plane, measured in the NIR regime in bands $I$ through $M$ (out to $\sim$5 \um). 
Again, although this law has \Rv dependence in the UV and optical, it is essentially fixed in the NIR.

\citet{fm07} use 328 near-main-sequence OB stars to revisit the UV through NIR extinction law,
but using the synthetic stellar model method \citep{Fitzpatrick2005} rather than the empirical pair method used in previous studies.
For $>$1 \um, however, they just assume $\betamath=1.84$ from \citet{Martin1990}, a value in very close agreement with our result.
\citet{Martin1990} investigate both diffuse and highly reddened sightlines. The diffuse data come from previous compilations \citep{Savage1979,Whittet1988a} of many sightlines generally in the Galactic plane, and the highly reddened sightlines are $\rho$ Oph and the Tr 14/16 OB stars. 
The \citet{Martin1990} study finds their data to be consistent with a single-\bnir universal curve between $I$ and at least 5 \um for both the diffuse and highly reddened regimes. 

In the most recent update to their work, \citet{Fitzpatrick2019} produce a new UV through NIR parameterization based on a sample of 72 OB stars. They have spectrophotometric data up to 1 \um, beyond which 2MASS photometry is used. The extinction curves are derived using synthetic stellar models, now in the NIR as well.
In that method, the observed SEDs of reddened near-main-sequence B (and late O) stars are modeled in order to simultaneously derive estimates of the physical properties of the star along with the shape of its extinction curve. \citet{Fitzpatrick2019} derive and present their extinction curves as a series of finely spaced points in the region where they have spectra, out to 1 \um. Beyond that, they present one point for each 2MASS band (\textit{JHK}), and a point representing $\lambda=\infty$. One can then interpolate to gain a continuous extinction curve. In the NIR, we find that their curve can be approximated by a power-law with $\betamath=1.86$ at \Rvstd, again in excellent agreement with our result.

To conclude, two of the earlier standard laws \citep[\ctccm;][]{Fitzpatrick1999} featured shallower NIR laws ($\betamath\approx 1.6$), whereas the two more recent laws feature somewhat steeper laws ($\betamath\approx 1.9$). Our results agree with the latter two.

\vfill\null
\subsubsection{NIR studies of moderately dusty sightlines}

In the intervening period between the \citet{Rieke1985} study and the review on dust grains by \citet{Draine2003}, several different values for \bnir were found, all larger than 1.61, with an upper bound around 1.8 \citep[e.g.,][]{Martin1990}. Since the turn of the century, derived values of \bnir have generally crept even higher. However, relatively few NIR studies have focused their attention toward diffuse regions of the Galaxy; most investigate the Galactic Center or dense clouds (such as the Coalsack). Those studies tend to find large values of \bnir (see Section \ref{subsubsec:beta_vs_dust}). Even so, some recent studies find diffuse sightlines to have relatively steep NIR slopes: 2.14 \citep{Stead2009}, 1.95 \citep{Wang2014}, and 2.07 \citep{Wang2019}. 

The \citet{Wang2019} study is perhaps the most similar in methods to ours in the literature to date. They select 61,111 red clump stars along diffuse Galactic sightlines, collecting photometry from a plethora of surveys from the optical to the MIR (including SDSS and 2MASS). To derive intrinsic colors, they employ stellar parameters from APOGEE \citep{Eisenstein2011} in a similar fashion to our method in Section \ref{sec:methods}. They include a second-order term in \teff, but not \logg. Their $\betamath=2.07$ is higher than ours with a small uncertainty (0.03), making it in disagreement with us.

\citet{Indebetouw2005} presents a bit of an outlier as far as recent diffuse \bnir: 1.65. However, they do not explicitly compute this value themselves; rather, it is quoted in other papers that way as derived from color excess ratios. \citet{Stead2009} present the fact that determining \bnir in this way depends sensitively on the wavelengths chosen for the filters. Using isophotal wavelengths, they find $\betamath=1.81$ for \citet{Indebetouw2005}; using effective wavelengths instead, they find $\betamath=2.05$. It is possible that this effect impacts \bnir values reported for other studies which calculate color excess ratios based entirely within \textit{JHK}. \citet{Stead2009} suggest that the use of isophotal wavelengths could be the reason for shallower NIR slopes in earlier studies. \cite{Fritz2011}, however, find that the use of effective wavelengths actually slightly overestimates \bnir. In our study, since we do not use color excess ratios to determine \bnir, we are much less sensitive to the choice of wavelength for \textit{JHK}.
Notably, \citet{Wang2019} also derive \bnir using color excess ratios, so they may be susceptible to similar issues.

While not large in sample size, the recent study of \citet{Decleir2022} uses very detailed spectrophotometric data that allows precise determination of the NIR extinction law. The unweighted average of their 13 stars yields $\betamath=1.77$, supporting the canonical value.

\subsubsection{NIR Slope vs. the Dust Column} \label{subsubsec:beta_vs_dust}

All of the "standard" laws discussed in Section \ref{subsubsec:standard} are derived from sightlines in the plane of the Galaxy. In that low-latitude regime, we find $\betamath=1.84$. 
Since in this study we are particularly interested in the less explored high-latitude regime, we determine a separate value for it: $\betamath=1.91$. The results between the two regimes are in better than $2\sigma$ agreement, so we find it appropriate to claim there are no differences in the character of the NIR power-law extinction between the low- and high-latitude regimes. In other words, \bnir values from studies based solely on sightlines confined to the Galactic plane should also apply at high latitude, and our agreement with canonical value supports that.


Many of the most recent studies of NIR extinction focus on sightlines toward the Galactic Center. These studies find preferentially larger values of \bnir than those which are focused away from the GC. This could be indicative of different NIR extinction behavior associated with "extreme" dust. In that region, $E(B-V) \sim 10$, far beyond what we probe in this study (even for our most reddened sightlines). Vexingly, though, there has not been uniform agreement across studies as to the value of \bnir even toward the GC alone. For example, \citet{Rieke1985} pull most of their sightlines from the GC, but they find $\betamath=1.61$, one of the smallest values in regular use. Modern studies are more consistent, with \bnir values ranging from $\sim$2 to >2.5 in that direction, depending on the study. Although the methods for constructing extinction curves are relatively well-understood, it would appear that something still lies hidden in the process which leads to a wide dispersion of results. 

There are a few studies which investigate both the GC and other Galactic plane regions together, thus standardizing technique and allowing for an internally consistent comparison between NIR extinction at different Galactic longitudes. One of these, \citet{Zasowski2009}, uses 2MASS NIR and \textit{Spitzer} MIR photometry to investigate the IR extinction law across 150$^\circ$ of longitude in the Galactic plane. 
They report their results in terms of color excess ratios, which can be converted to \bnir by assuming the underlying dependence is a power law. Thus their $E(H-5.8)/E(H-K)$ values yield $\betamath=2.2$ for $|l|<25^\circ$ and $\betamath=1.8$ for $|l|\sim70^\circ$. The latter result stands in excellent agreement with our results, whereas the former result agrees equally well with the most recent estimate for the GC inner $3 \times 3$ deg$^2$ extinction law from \citet[][also $\betamath=2.2$]{Sanders2022}.
Although the dependence of color excess ratio on longitude in \citet{Zasowski2009} unarguably leads to the conclusion of a different law near and away from the central region of the Galaxy, the more recent study of
\citet{MaizApellaniz2020}
find the same $\betamath=2.26$ near the Galactic Center and other regions of the plane.

While there is evidence that the slope of the NIR extinction curve takes a canonical value everywhere in the sky except near the center of the Galaxy, some studies find steeper values elsewhere in the plane, not allowing us to reach a final conclusion as to where the change between steep and less steep laws occurs.

\subsubsection{NIR Slope vs. {\normalfont \Rv}}

In Section \ref{subsec:bnirRv}, we show that the apparent correlation we find between \bnir and \Rv can be explained by covariant uncertainties in those parameters, and that the intrinsic variation in \bnir has a standard deviation of only about 0.1. In other words, we confirm a universal NIR law in the sense that there are no \Rv-dependent trends as there are in the UV and optical regimes. 

The strongest evidence for an \Rv dependence so far was found by \citet{Fitzpatrick2009}. They find that, for a selection of OB stars, \bnir does vary with \Rv. The orange points in the right panel of Figure \ref{fig:betaNIR_Rv} are taken from that study, with associated fit to the trend from \citet[their Fig. 4D]{Salim2020}. 
The result from \citet{Fitzpatrick2009} is in stark contrast to those presented/assumed in previous parameterizations of the optical+NIR extinction curve \cite[e.g., \ctccm,][]{Fitzpatrick1999, fm07}, which provide extinction laws dependent on the single parameter \Rv, \textit{except} in the NIR, where the curve becomes invariant. Though only 14 sightlines are studied in \cite{Fitzpatrick2009}, they show a relatively robust correlation between \bnir and \Rv. They suggest that
it would be ``surprising'' for extinction to follow a constant power-law everywhere across a large wavelength range.
\citet{Fitzpatrick2009} attribute the ability to see the correlation to a large (intrinsic) range of \Rv values in their sample. 

\citet{Fitzpatrick2019} provide an updated \Rv-dependent Milky Way extinction curve in which \bnir is no longer assumed to be mostly constant as was the case in \ctccm and \citet{Fitzpatrick1999}. The resulting relationship between \bnir and \Rv (shown in the right panel of Figure \ref{fig:betaNIR_Rv}) is nonetheless quite a bit shallower than the one from \citet{Fitzpatrick2009}, only spanning a range of $\sim$0.5 in \bnir on $2<R_V<6$. Such a weak dependence may not be detectable in our analysis, 
where we find the \Rv range is quite small.
This relationship and its origin are not specifically discussed in the study, but are inherent in the presented \Rv-dependent curves. 


In a recent study, \citet{Decleir2022} use NIR spectra to construct extinction curves along 15 sightlines via the pair method. Their sample is similar to that of \citet{Gordon2021},  including 5 of the same stars. The sightlines are quite dust-heavy compared to ours, with two dense enough to show an ice extinction feature. \citet{Decleir2022} use IRTF/SpeX spectra covering the NIR and find a moderate negative correlation between \Rv and \bnir ($\rho=-0.68$).
In the right panel of Figure \ref{fig:betaNIR_Rv}, we show the diffuse sightlines from \citet{Decleir2022} in purple (excluding two with large \Rv uncertainties). They follow a similar relation to the \citet[][orange]{Fitzpatrick2009}.

Analysis of our sample reveals that the \Rv range is quite small, so any dependence of \bnir on \Rv would be relatively subtle. Nonetheless we test, again using the mock analysis, different degrees of linear correlation ($\rho$) between \bnir and \Rv. We find that $\rho$ departing from zero yields a poorer match between the mock and actual measurements. From this we conclude that, for the general, low-latitude sightlines we have in our sample of individual curves, there is no significant correlation between \bnir and \Rv. Our findings contrast the studies discussed above that use "representative" samples, which select sightlines with a wide range of values in parameter space (especially \Rv). Within these samples, it would appear that the inclusion of sightlines with $R_V$ significantly beyond 4 may lend itself to the finding of a correlation between \Rv and \bnir.

There have been a few studies that explore the relationship between the NIR extinction law and \Rv using statistical samples. 
One such study, \citet{Schlafly2016}, derived the extinction law in the Galactic plane using stars with APOGEE spectra and optical (Pan-STARRS1) and NIR (2MASS and WISE) photometry. They find no correlation between $E(y-J)/E(J-H)$ or $E(J-H)/E(H-K)$ and \Rv (their Figure 12, bottom panel). Like us, they find \Rv to have a small range ($\sigma(R_V)=0.18$), so sightlines that may potentially show a non-universal NIR law are uncommon. A similar conclusion of a lack of dependence between the NIR extinction curve slope and \Rv can be gleaned from the \citet{Nataf2016} study of extinction near the GC. Their Figure 8 shows that $E(J-K_s)/E(I-J)$ does not depend on $R_I=A_I/E(V-I)$, which should correlate with \Rv.

The landscape of results pertaining to a dependence of NIR slope on \Rv is confusing given that some studies do and others do not find a dependence. Of the two that do \citep{Fitzpatrick2009, Decleir2022}, the key seems to be having sightlines with $R_V>4$, which seem to be rare in general samples of stars like ours.
One the other hand, according to \citet{Martin1990}, the $\rho$ Oph molecular cloud has $R_V=4.3$, but its NIR slope ($1.85\pm0.09$) agrees with the \bnir found in the same study for diffuse \Rvstd sightlines ($1.84\pm0.03$). It should be pointed out, however, that $\rho$ Oph stars have $1.1<E(B-V)<2.9$, where as the two stars with $R_V>4$ in \citet{Decleir2022} have $E(B-V)\sim 0.5$. Similarly, the four stars with $R_V>4$ in \citet{Fitzpatrick2009} also have lower reddening ($0.5<E(B-V)<0.9$) than $\rho$ Oph. Could it be that in the case of $\rho$ Oph, high \Rv coupled with high dust column somehow conspire to yield a "normal" \bnir?

Both \citet{Fitzpatrick2009} and \citet{Decleir2022} are based on very small samples and \citet{Fitzpatrick2009} may include some problematic stars \citep{MaizApellaniz2018}. The hope is that the sample of stars for which \bnir and \Rv can be determined well enough to not be affected by covariances will be significantly expanded in the future so that the question can be definitively answered. 

\subsection{$\sigma(R_V)$} \label{subsec:disc_Rv}

Much effort has been put into the determination of \Rv in the literature since \ctccm codified it as the single value upon which the extinction curve in the UV and optical relies. The value determined in that study, \Rvstd, is the one still generally applied today. Many subsequent studies \citep[e.g.,][]{Fitzpatrick1999, fm07, Schlafly2011} have found general agreement with this value, either by direct measurement of \Rv or by comparing their extinction values to those predicted by curves of different \Rv. As noted above, we adopt \Rvstd for the generation of the mock stars in our analysis. We find no evidence to contradict that value from our data. 

Of greater interest to our study is the intrinsic spread in \Rv, \sigRv. While \Rvstd is characteristic of the Milky Way \textit{average} extinction curve, it is important to know the chance that any given sightline may vary from that value significantly. We derive the spread using a mock analysis in two different ways (see Section \ref{subsec:bnirRv}), yielding the similar results $\sigma=0.28$ and $0.24$.
\citet{fm07} find $\sigma=0.27$ from their sample of 328 extinction curves along sightlines to OB stars. \citet{Schlafly2016} find a narrower spread with $\sigma=0.18$, using 37,000 stars from the APOGEE survey largely confined to the Galactic plane \citep[roughly the same width as found earlier by][]{Schlafly2010}. The characteristics of the distributions found by \citet{fm07} and \citet{Schlafly2016} are somewhat different. \citet{fm07} found quite a strong high-\Rv tail to their distribution, such that the Gaussian which yielded their spread was only fit to a core region around $R_V=3.0$; the tail dragged the mean of the entire sample up to $R_V=3.22$. \citet{Schlafly2016} find an almost symmetric Gaussian distribution about $R_V=3.32$, with a very small tail; they posit that the difference could be due to the different populations of stars probed (OB stars vs. background giants). 
They further note that many high-\Rv stars in \citet{fm07} are relatively poor fits, and claim that there may be an even smaller number of high-\Rv sightlines in reality (i.e., essentially no tail). We assumed a Gaussian distribution for intrinsic \Rv in our mock analysis partially due to this finding. In the future, it may be useful to carry out a more detailed analysis which allows for tailed and/or non-Gaussian distributions of intrinsic \Rv.


\vfill\null
\subsection{Cepheid Calibration} \label{subsec:disc_H0}

In Section \ref{subsec:H0}, we find a new value for the NIR extinction parameter used to correct Cepheid variable magnitudes: \RHuserr. Our value is somewhat lower than the \citet{Fitzpatrick1999} \RHstd used in \citet{Riess2016,Riess2021,Riess2022}. 
In \citet{Riess2022} an additional analysis was performed to determine, rather than assume, the value of \RH by searching for the value that improves the overall fit the best, yielding $R_H=0.363\pm0.038$ for the MW Cepheid sample. This value comes somewhat closer to what we find (but is still consistent with the \citet{Fitzpatrick1999} value as well).

In the right panel of Figure \ref{fig:RHRv}, we show the relation between \RH and \Rv which comes directly from the literature extinction laws \ctccm (purple), \citet[][orange]{Fitzpatrick1999}, and \citet[][green]{Fitzpatrick2019}. For \citet{Fitzpatrick1999}, there is almost no dependence of \RH on \Rv. 
We also show a cyan point at $R_H=0.336$ for \citet{fm07}, in which \Rvstd. At $R_V=3.1$, \citet{Fitzpatrick2019} yields $R_H=0.411$.
Our derived intrinsic values (gray contours) are in general agreement with these standard curves in the relevant \Rv range. 

\citet{Riess2022} find the relation 
$\Delta (5 \log H_0) = 0.08\Delta R_H$, and therefore $\Delta H_0/H_0 = 0.04 \Delta R_H$.
Adjusting from their baseline value of $H_0 = 73.04$ km s$^{-1}$ Mpc$^{-1}$, our \RHus would yield $H_0 = 73.16$ km s$^{-1}$ Mpc$^{-1}$, 
a change of $\sim$0.1$\sigma$.

We note that the recent study \citet{Mortsell2022} finds a very strong change in the fitted value of \hub with increasingly stringent cuts on the reddening \EVI of the Cepheids included in their sample. Given that we find no correlation between \RH and \Euz (or by extension \EVI), our results would not support such a change in the final \hub value.


\section{Conclusions} \label{sec:conclusions}
In this work, we used $\sim$50,000 stars with SDSS spectra and SDSS+2MASS photometry to investigate the nature of the Milky Way extinction curve in the near-infrared (out to $K$ band). We examine whether the curve is universal or varies with the amount of dust or with \Rv.
\begin{enumerate}
    \item We report an overall mean power-law slope of the near-infrared extinction curve $\betamath=1.85 \pm 0.01$. We recommend this \bnir for correcting the magnitudes of extragalactic objects, MW Cepheids, and other objects with low to moderate reddening ($0.02<E(B-V)\lesssim1$).

    \item We find no correlation between \bnir and the dust column over a wide range of dust column densities, from very low ones at high latitudes ($E(B-V) \sim 0.02$) to moderately high ones at lower latitudes ($E(B-V) \sim 0.8$). However, we do not explore highly reddened regions such as the Galactic Center, where some studies find steeper curves ($\betamath \gtrsim 2$).

    \item Our \bnir confirms the NIR law used in \citet{fm07} and \citet{Fitzpatrick2019}. It is however steeper than in either \citet{CCM} or \citet{Fitzpatrick1999} ($\betamath \sim 1.6$), so we do not recommend those two for the NIR. We do not determine which standard law may be favored at shorter wavelengths. We propose replacing the $\lambda>7500$ \AA\ portion of whichever law one uses with 
    \begin{equation*}
        A_\lambda = A_{7500}\left(\frac{\lambda}{7500 \text{ \AA}}\right)^{-1.85}.
    \end{equation*} 

    \item We find no strong evidence for a correlation between \Rv and the dust column, although there is weak evidence for a rise beyond \Rvstd below $E(B-V)\approx0.1$.

    \item There is an apparent correlation between \bnir and \Rv (i.e., between NIR and optical slopes, since $R_V^{-1}=A_B/A_V-1$), which we entirely explain (using a mock analysis) as arising from covariant uncertainties in the two parameters. We conclude that an intrinsic correlation between \bnir and \Rv must be small or nonexistent in our sightlines. Some recent studies that do find a correlation include sightlines with $R_V>4$, which seem very rare in our general sample.

    \item While the apparent star-to-star scatter in \bnir is quite large, we find (again using the mock analysis) that this is the result of observational uncertainties, and that the intrinsic star-to-star scatter in \bnir is relatively small ($\sigma(\betamath)=0.13$). 

    \item With a similar mock analysis, we derive a relatively small intrinsic spread in \Rv of $\sigma(R_V)=0.24$, confirming the results of \citet{Schlafly2016} 
    and other large-sample studies.

    \item We derive a new value for the reddening parameter used in Cepheid calibrations in the measurement of \hub: \RHuserr. Like \bnir, \RH does not appear to intrinsically depend on the amount of dust or on \Rv. 
    
\end{enumerate}

While we have attempted to not only present our results but also arrive at a consistent picture, the landscape of literature results regarding NIR extinction is somewhat difficult to navigate. One thing that may help in the future is to consider using mock analysis and forward modeling to test whether any correlation between extinction parameters is robust.

\begin{acknowledgments}
The authors would like to acknowledge the use of several Python packages: Scipy \citep{scipy}, \texttt{scikit-learn} \citep{scikit-learn}, \texttt{extinction} \citep{extinction}, and \texttt{dust\_extinction} \citep{dust-extinction}. We also acknowledge the use of the ApJ template developed by Cubillos and Harrington (\href{https://github.com/pcubillos/ApJtemplate}{https://github.com/pcubillos/ApJtemplate}) in the drafting of this manuscript.
\end{acknowledgments}



\end{document}